%% file: Paper.tex
\documentclass[a4paper, 10pt]{article}

\usepackage{jheppub}

\usepackage{amsthm}

\usepackage[numbers,sort&compress]{natbib}
\usepackage{dsfont}
\usepackage{slashed}
\usepackage{color}
\usepackage{empheq}

\usepackage{adjustbox}

\usepackage{bibgerm}

\usepackage{soul}

\usepackage[english]{babel}

\usepackage[utf8x]{inputenc}

\usepackage{graphicx,epsfig}
\usepackage{pstricks}

\usepackage{tikz}

\usepackage{bashful}

\usepackage{subfigure}

\usepackage{setspace}

\usepackage{booktabs}

\usepackage{float}

\usepackage{nextpage}

\usepackage{pdfpages}

\usepackage{hypcap}

\usepackage[small,bf]{caption}

\usepackage{longtable}

\usepackage{fancyhdr}

\usepackage[makeroom]{cancel}

\usepackage{microtype}

\usepackage{xparse}
\usepackage{etoolbox}

\usepackage[subfigure]{tocloft}

\usepackage{multirow}

\usepackage[all]{nowidow}

\renewcommand{\Re}{\mathrm{Re}}
\renewcommand{\Im}{\mathrm{Im}}

\newcommand{\inel}{\mathrm{in}}
\newcommand{\<}{\langle}
\renewcommand{\>}{\rangle}

\renewcommand{\O}{\mathcal{O}}

\newcommand{\nn}{\nonumber\\}

\newcommand{\EL}{E\L{}}
\newcommand{\GeV}{\,\text{GeV}}
\newcommand{\MeV}{\,\text{MeV}}
\newcommand{\fm}{\,\text{fm}}
\newcommand{\sthr}{s_\mathrm{thr}}

\newcommand{\XXint}[3]{{\setbox0=\hbox{$#1{#2#3}{\int}$}
\vcenter{\hbox{$#2#3$ }}\kern-.65\wd0}}

\definecolor{darkgreen}{rgb}{0,0.5,0}
\definecolor{darkblue}{rgb}{0,0,0.5}
\definecolor{darkred}{rgb}{0.5,0,0}
\definecolor{beige}{rgb}{0.7,0.4,0.3}

\makeatletter
  \def\my@tag@font{\normalsize}
  \def\maketag@@@#1{\hbox{\m@th\normalfont\my@tag@font#1}}
  \let\amsmath@eqref\eqref
  \renewcommand\eqref[1]{{\let\my@tag@font\relax\amsmath@eqref{#1}}}
\makeatother

\makeatletter
\renewcommand\paragraph{\@startsection{paragraph}{4}{\z@}%
  {-3.25ex\@plus -1ex \@minus -.2ex}%
  {1.5ex \@plus .2ex}%
  {\normalfont\normalsize\bfseries}}
\makeatother

\cftsetindents{section}{0em}{2em}
\cftsetindents{subsection}{2em}{3em}
\cftsetindents{subsubsection}{5em}{4em}

\allowdisplaybreaks[1]

\preprint{
\mbox{}\hfill{} PSI-PR-24-21 \\
\mbox{}\hfill{} ZU-TH 03/25
}

\title{\boldmath Dispersive analysis of the pion vector form factor without zeros}

\author[a,b,c]{Thomas P. Leplumey,}

\author[b,d]{Peter Stoffer}

\emailAdd{leplumey@llr.in2p3.fr}
\emailAdd{stoffer@physik.uzh.ch}

\affiliation[a]{Institute for Particle Physics and Astrophysics, ETH Z\"urich, 8093 Z\"urich, Switzerland}
\affiliation[b]{Physik-Institut, Universit\"at Z\"urich, Winterthurerstrasse 190, 8057 Z\"urich, Switzerland}
\affiliation[c]{Ecole Polytechnique, IN2P3-CNRS, Laboratoire Leprince-Ringuet, F-91128 Palaiseau, France}
\affiliation[d]{PSI Center for Neutron and Muon Sciences, 5232 Villigen PSI, Switzerland}

\abstract{
We perform an updated analysis of $e^+e^-\to\pi^+\pi^-$ cross-section data using a dispersive representation of the pion vector form factor. We show that the available data are compatible with the assumption that the form factor is free of complex zeros and that under this assumption the largest systematic uncertainty in a previous analysis can be eliminated. We investigate both a constrained Omn\`es representation as well as a hybrid phase-modulus representation and we quantify the discrepancies in the hadronic vacuum polarization contribution to the anomalous magnetic moment of the muon based on different $e^+e^-$ data sets. We find that the dispersive constraints exacerbate these discrepancies. Together with the assumption of the absence of zeros, the pion charge radius becomes a useful observable to discriminate between the different data sets. This provides an opportunity for future improved lattice-QCD determinations to probe the discrepancies independently of full computations of hadronic vacuum polarization. We also reevaluate the two-pion contribution to Euclidean windows and we observe that systematic discrepancies between the data sets persist even at very long distances.
}

\numberwithin{equation}{section}

\begin{document}

	\maketitle


	\input{sections/Introduction}

	\input{sections/ConstrainedFits}

	\input{sections/ModulusRepresentation}

	\input{sections/Fits}

	\input{sections/Conclusions}

	
	\section*{Acknowledgements}
	\addcontentsline{toc}{section}{\numberline{}Acknowledgements}

	PS thanks G.~Colangelo and M.~Hoferichter for long-standing collaboration on the subject, which the present work draws on. 
	We thank B.~Ananthanarayan, G.~Colangelo, A.~X.~El-Khadra, M.~Hoferichter, B.~Kubis, B.~Malaescu, E.~Ruiz~Arriola, and P.~S\'anchez-Puertas for useful discussions and M.~Hoferichter for helpful comments on the manuscript.
	Financial support by the Swiss National Science Foundation (Project No.~PCEFP2\_194272) is gratefully acknowledged.

	
	\appendix

	\input{sections/Hybrid}

	\input{sections/Radius}

	\clearpage

	\addcontentsline{toc}{section}{\numberline{}References}
	\bibliographystyle{utphysmod}
	\bibliography{Literature}
	
\end{document}

%% file: sections/Introduction.tex

\section{Introduction}

The anomalous magnetic moment of the muon $a_\mu$ has received a lot of attention over the past years due to the discrepancy between the experimental value and the theoretical prediction within the Standard Model (SM). While the new measurements at Fermilab E989~\cite{Muong-2:2021ojo,Muong-2:2023cdq,Muong-2:2024hpx} confirmed and improved over previous results from E821~\cite{Muong-2:2006rrc}, the situation on the theory side has become much more puzzling since the first White Paper (WP) in 2020~\cite{Aoyama:2020ynm}, see Ref.~\cite{Colangelo:2022jxc} for a recent summary. The challenge lies in a reliable evaluation of the non-perturbative hadronic contributions to $a_\mu$, in particular hadronic vacuum polarization (HVP) and hadronic light-by-light (HLbL), which completely dominate the theory uncertainty. The two main approaches to evaluate the HVP and HLbL contributions are on the one hand data-driven methods, making use of dispersion relations and experimental input, and on the other hand lattice QCD. In the case of HLbL, lattice-QCD evaluations~\cite{Blum:2019ugy,Chao:2021tvp,Blum:2023vlm,Fodor:2024jyn} agree with the data-driven result reported in the WP~\cite{Aoyama:2020ynm,Melnikov:2003xd,Pauk:2014rta,Colangelo:2014dfa,Colangelo:2014qya,Colangelo:2014pva,Colangelo:2015ama,Danilkin:2016hnh,Masjuan:2017tvw,Colangelo:2017qdm,Colangelo:2017fiz,Jegerlehner:2017gek,Knecht:2018sci,Hoferichter:2018kwz,Gerardin:2019vio,Hoferichter:2019nlq,Bijnens:2019ghy,Roig:2019reh,Eichmann:2019bqf,Colangelo:2019lpu,Colangelo:2019uex}. On the data-driven side, improvements since the release of the WP~\cite{Ludtke:2020moa,Bijnens:2020xnl,Bijnens:2021jqo,Danilkin:2021icn,Colangelo:2021nkr,Stamen:2022uqh,Bijnens:2022itw,Ludtke:2023hvz,Stoffer:2023gba,Hoferichter:2024fsj,Ludtke:2024ase,Holz:2024lom,Holz:2024diw} have made it possible to reach the precision goal of 10\% on the HLbL contribution~\cite{Hoferichter:2024bae,Hoferichter:2024vbu}.

For HVP, the situation is much more convoluted. The tension between the BMWc lattice-QCD computation of Ref.~\cite{Borsanyi:2020mff} with data-driven results~\cite{Davier:2017zfy,Keshavarzi:2018mgv,Colangelo:2018mtw,Hoferichter:2019mqg,Davier:2019can,Keshavarzi:2019abf,Hoid:2020xjs,Colangelo:2021moe,Colangelo:2022prz,Hoferichter:2023sli} has been confirmed in the last years by further complete or partial lattice results~\cite{Ce:2022kxy,ExtendedTwistedMass:2022jpw,FermilabLatticeHPQCD:2023jof,RBC:2023pvn,Boccaletti:2024guq,RBC:2024fic,Djukanovic:2024cmq,Bazavov:2024eou}. Regarding data input, the CMD-3 measurement~\cite{CMD-3:2023rfe,CMD-3:2023alj} is in conflict with all previous high-statistics $e^+e^-$ two-pion cross-section results, but agrees with both the lattice evaluations and a scenario without New Physics in $a_\mu$. Clarifying the origin of these discrepancies is of utmost importance, in order to arrive at a consolidated SM prediction for $a_\mu$ that matches the experimental uncertainty and enables a correct and robust interpretation of the upcoming final result of the Fermilab $g-2$ experiment.

Even before the CMD-3 result, the origin of the tension between the lattice and dispersive HVP evaluations was identified to be located in the low-energy region below $2\GeV$, with important effects below $1\GeV$~\cite{Crivellin:2020zul,Keshavarzi:2020bfy,Malaescu:2020zuc,Colangelo:2020lcg}. With the CMD-3 result, it is now evident that a better understanding of the two-pion channel at low energies is crucially needed. In particular, radiative corrections are under scrutiny~\cite{Campanario:2019mjh,Monnard:2021pvm,Ignatov:2022iou,Colangelo:2022lzg,BaBar:2023xiy,Budassi:2024whw}, including a community effort to assess and improve Monte Carlo generators~\cite{Abbiendi:2022liz,Aliberti:2024fpq}, but although a class of structure-dependent contributions to the radiative-return process has been identified as a potential source of uncontrolled model dependences~\cite{Aliberti:2024fpq}, at present it is not clear if such effects could be of the size of the observed discrepancies.

The dispersive evaluation of HVP is based on the principles of unitarity and analyticity. The two-pion channel is directly related to the vector form factor (VFF) of the pion, and the very same fundamental principles used for HVP can be applied to this sub-process. Dispersive analyses of the VFF~\cite{Leutwyler:2002hm,Colangelo:2003yw,Ananthanarayan:2011xt,Hanhart:2012wi,Ananthanarayan:2012tt,Ananthanarayan:2013zua,Ananthanarayan:2016mns,Ananthanarayan:2017efc,Ananthanarayan:2018nyx,Colangelo:2018mtw,Ananthanarayan:2020vum,Colangelo:2020lcg,Colangelo:2022prz,Stoffer:2023gba,Simula:2023ujs,RuizArriola:2024gwb} offer the opportunity to scrutinize the cross-section data, they allow for an analytic continuation to the space-like region, and they provide a link to further observables, such as the pion charge radius. In the present work, we extend the dispersive analysis of Refs.~\cite{Leutwyler:2002hm,Colangelo:2003yw,Colangelo:2018mtw,Colangelo:2020lcg,Colangelo:2022prz,Stoffer:2023gba}, which incorporates the constraints of unitarity and analyticity by relying on the solution of the Roy equations for the $\pi\pi$ phase-shift input. However, the dispersive parametrization introduces systematic theory uncertainties, which dominated the uncertainties of the fit results of Refs.~\cite{Colangelo:2018mtw,Colangelo:2020lcg,Colangelo:2022prz,Stoffer:2023gba} in particular for the pion charge radius, diluting to some extent the power of the dispersive constraints. Here, we show that the largest systematic uncertainty in the dispersive analysis is related to the appearance of complex zeros in the VFF. It has been argued in the past that the VFF should be free from zeros~\cite{Leutwyler:2002hm}. We show that by imposing the absence of zeros as a constraint in the dispersive fits, the systematic uncertainties are drastically reduced, whereas the fit quality remains largely unaffected. This has interesting consequences: the reduction of the uncertainties in $a_\mu$ compared to a direct integration of cross-section data~\cite{Davier:2017zfy,Keshavarzi:2018mgv,Keshavarzi:2019abf,Davier:2019can} becomes more powerful beyond the region of very low energies. Consequently, the discrepancies in $a_\mu$ based on input from different $e^+e^-$ data sets get enhanced. These discrepancies persist in the extrapolation to low energies, in particular we find tensions even in very-long-distance Euclidean windows, but also in the pion charge radius, which becomes an interesting probe to discriminate between the data sets. This offers the opportunity for future lattice-QCD evaluations with reduced uncertainties to test the discrepancies with an independent observable.

The article is structured as follows. In Sect.~\ref{sec:Implementation}, we review the Omn\`es representation used in our dispersive approach, we explain how we impose the absence of zeros, and we adopt an improved treatment of the elasticity input parameter, which has been responsible for the second-largest systematic uncertainty. In Sect.~\ref{sec:HighEnergyContinuation}, we employ a hybrid phase-modulus dispersion relation in order to extend our representation beyond $1\GeV$, which leads to results compatible with the Omn\`es representation in the low-energy region. In Sect.~\ref{sec:Fits}, we present our fit results, whereas in Sect.~\ref{sec:Results} we discuss the consequences for $a_\mu$, the Euclidean windows, and the pion charge radius, before concluding in Sect.~\ref{sec:Conclusions}.

%% file: sections/ConstrainedFits.tex

\section{Omn\`es representation}
\label{sec:Implementation}

\subsection{Dispersive representation}
\label{sec:DispersiveRepresentation}

In the following, we briefly review the dispersive representation from Refs.~\cite{Leutwyler:2002hm,Colangelo:2003yw,Colangelo:2018mtw,Colangelo:2020lcg,Colangelo:2022prz,Stoffer:2023gba}. The pion VFF can be decomposed into three factors,
\begin{equation}
	\label{eq:VFFOmnesRepresentation}
	F_\pi^V(s) = \Omega_1^1(s) \times G_\omega(s) \times G_\inel(s) \,,
\end{equation}
where the Omn\`es function~\cite{Muskhelishvili:1953,Omnes:1958hv}
\begin{equation}
	\Omega_1^1(s)=\exp\left\{\frac{s}{\pi}\int_{4M_\pi^2}^\infty ds'\frac{\delta_1^1(s')}{s'(s'-s)}\right\}
\end{equation}
describes the effect of two-pion intermediate states in terms of the isospin $I = 1$ elastic $\pi\pi$ $P$-wave phase shift $\delta_1^1(s)$ in the isospin limit. The phase shift itself is constrained to fulfill the $\pi\pi$ Roy equations~\cite{Roy:1971tc}, which have been solved to high precision in Refs.~\cite{Ananthanarayan:2000ht,Garcia-Martin:2011iqs,Caprini:2011ky,Pelaez:2024uav}. We rely on the solutions of Refs.~\cite{Ananthanarayan:2000ht,Caprini:2011ky}, which are parametrized in terms of the $I=1$ $P$-wave phase shift at $s_0=(0.8\GeV)^2$ and $s_1=(1.15\GeV)^2$. For a phase reaching asymptotically $\delta_1^1(s) \asymp \pi$, the Omn\`es function behaves as $\Omega_1^1(s) \asymp s^{-1}$.

The second factor $G_\omega(s)$ takes into account isospin-breaking effects, which are not negligible if resonantly enhanced. As in Refs.~\cite{Colangelo:2018mtw,Colangelo:2022prz,Stoffer:2023gba}, we use a dispersively improved representation of the narrow $\omega$ resonance, which ensures the absence of unphysical sub-threshold imaginary parts and the correct threshold behavior. The impact of additional radiative channels (mainly $\pi^0\gamma$) can be accounted for by an effective complex phase in the $\rho-\omega$ coupling $\epsilon_\omega$, leading to~\cite{Colangelo:2022prz}
\begin{align}
	\label{eq:G_omega}
	G_V(s) = 1
		&+\frac{s}{\pi}\int_{9M_\pi^2}^\infty ds'\frac{\Re\,\epsilon_V}{s'(s'-s)}\Im\left[\frac{s'}{\left(M_V-\frac{i}{2}\Gamma_V\right)^2-s'}\right]\left(\frac{1-\frac{9M_\pi^2}{s'}}{1-\frac{9M_\pi^2}{M_V^2}}\right)^4 \nn
		&+\frac{s}{\pi}\int_{M_{\pi^0}^2}^\infty ds'\frac{\Im\,\epsilon_V}{s'(s'-s)}\Re\left[\frac{s'}{\left(M_V-\frac{i}{2}\Gamma_V\right)^2-s'}\right]\left(\frac{1-\frac{M_{\pi^0}^2}{s'}}{1-\frac{M_{\pi^0}^2}{M_V^2}}\right)^3 \,,
\end{align}
with $V=\omega$, where the part proportional to $\Re\,\epsilon_\omega$ describes the effects of the $3\pi$ channel, while the part proportional to $\Im\,\epsilon_\omega$ accounts for the effects of the radiative channels starting with $\pi^0\gamma$. We take the omega width $\Gamma_\omega=8.71(3)\MeV$ as input from an analysis of the three-pion channel~\cite{Hoferichter:2023bjm}. Fitting the $\omega$ width instead of keeping it as a fixed input does not improve the fit quality~\cite{Colangelo:2018mtw}. The main effect would be a strong correlation with the complex phase of the mixing parameter, $\delta_\epsilon = \arg(\epsilon_\omega)$. A similar correlation between $\delta_\epsilon$ and $M_\omega$ has been observed before~\cite{Colangelo:2022prz}.

The third factor $G_\inel(s)$ describes the effects from all other inelastic channels. In Refs.~\cite{Colangelo:2018mtw,Colangelo:2022prz,Stoffer:2023gba}, it is represented by a conformal polynomial
\begin{equation}
	\label{eq:Gin}
	G_\inel(s) = P_N(z(s)) = 1+\sum_{k=1}^Nc_k(z^k(s)-z^k(0)) \,,
\end{equation}
where the conformal variable is
\begin{equation}
	\label{eq:ConfZ}
	z(s)=\frac{\sqrt{s_\inel-s_c}-\sqrt{s_\inel-s}}{\sqrt{s_\inel-s_c}+\sqrt{s_\inel-s}} \,,
\end{equation}
and inelasticities become relevant only above the $\pi^0\omega$ threshold $s_\inel = (M_{\pi^0} + M_\omega)^2$. The point that gets mapped to the origin is varied in the range $s_c = -(0.5\ldots2)\GeV^2$. The requirement that the conformal polynomial reproduce $P$-wave threshold behavior, $\Im\,G_\inel(s)\sim(s-s_\inel)^{3/2}$, corresponds to $P_N'(1)=0$, i.e.,
\begin{equation}
	c_1 = -\sum_{k=2}^N k c_k \,,
\end{equation}
whereas the remaining $N-1$ parameters $c_k$ are kept free in the fit. This parametrization ensures the correct analytic structure, the normalization $G_\inel(0) = 1$, as well as the asymptotic behaviour $G_\inel(s) \asymp \text{const}$.
In addition, the inelastic phase of the form factor is constrained by the Eidelman-\L{}ukaszuk (\EL) bound~\cite{Eidelman:2003uh,Lukaszuk:1973jd}, which in this context takes the form~\cite{Colangelo:2018mtw}
\begin{equation}
	\label{eq:ELbound}
	\left|\arg\,G_\inel(s)\right|^2\leq\iota_1r\frac{(s-4M_\pi^2)^{3/2}(s-s_\inel)^{3/2}}{s_a^2} \,,
\end{equation}
with $s_a = (1\GeV)^2$ and $\iota_1 = 0.05(4)$. The \EL{} bound is implemented with a penalty in the $\chi^2$~\cite{Colangelo:2018mtw}.

\subsection{Zeros in the vector form factor}
\label{sec:Zeros}

In contrast to the Omn\`es function $\Omega_1^1(s)$ and $G_\omega(s)$, the factor $G_\inel(s)$ parametrizing inelastic effects in terms of the conformal polynomial~\eqref{eq:Gin} potentially develops zeros in the complex $s$ plane if $N\ge3$.

In Refs.~\cite{Colangelo:2018mtw,Colangelo:2022prz,Stoffer:2023gba}, the order $N$ of the conformal polynomial was varied in the range $N=2\ldots6$. The impact of this variation on the results was taken as a theoretical uncertainty, which turned out to dominate not only the entire systematics, but was also larger than the fit errors~\cite{Colangelo:2018mtw}. Even larger orders of the conformal polynomial were excluded because they led to an unphysical oscillating inelastic phase. Interestingly, in all fit results a rather large change can be observed when the order $N$ increases from $4$ to $5$. By explicitly inspecting the fit values for the coefficients of the conformal polynomial, one can verify that this sudden change coincides with the appearance of complex zeros in $G_\inel(s)$.

In Ref.~\cite{Leutwyler:2002hm}, arguments were presented why the VFF should not have complex zeros, based on the chiral expansion at low energies, the asymptotic behavior at high energies, the available data on the VFF, and a heuristic analogy with the hydrogen atom. Using dispersive methods, zeros have been excluded in a large region in the complex plane in Ref.~\cite{Ananthanarayan:2011xt} and more recently in Ref.~\cite{RuizArriola:2024gwb}.

Under the assumption that the VFF does not have complex zeros, the large variations obtained in Ref.~\cite{Colangelo:2018mtw} related to the order $N$ of the conformal polynomial overestimate the actual systematic uncertainty of the dispersive representation. Leaving a more thorough investigation of the possibility of zeros for future work~\cite{ZerosInPrep:2025}, in the following we discuss modifications of the dispersive representation that exclude the presence of complex zeros.

\subsection{Explicit zero-free parametrization}
\label{sec:ExplicitZeroFree}

The presence of zeros in the inelastic factor is most easily discussed in terms of the conformal variable $z = z(s)$. Excluding zeros in the VFF amounts to excluding zeros in the conformal polynomial $G_\inel(s) =: f(z(s))$ inside the unit disk of the complex $z$-plane. If a free fit prefers zeros inside the unit disk, their exclusion will typically push them to the unit circle. A zero at $z=1$, which represents the inelastic threshold $s=s_\inel$, is excluded from data. In order not to change the asymptotic behavior of the VFF
, we also exclude a zero at $z=-1$, which corresponds to $|s| = \infty$. Therefore, we consider the possibility of complex zeros on the unit circle, which need to come in pairs of complex conjugates.

A polynomial $f(z)$ with $2r$ roots on the unit circle with phases $\pm\varphi_k$, with $\varphi_k\in(0,\pi)$ can be factorized as
\begin{equation}
	f(z)=Q_{r,n}(z)=\prod_{k=1}^r\frac{1 - 2 z \cos\varphi_k + z^2}{1 - 2 z_0 \cos\varphi_k + z_0^2}P_n(z) \,,
\end{equation}
where $n=N-2r$, $z_0 = z(0)$, and $P_n$ is a polynomial parametrized as in Eq.~\eqref{eq:Gin}, assumed to have no zeros inside the unit disk. Due to
\begin{equation}
	\frac{Q_{r,n}'(1)}{Q_{r,n}(1)}=r+\frac{P_n'(1)}{P_n(1)} \,,
\end{equation}
the threshold-behavior constraint $f'(1)=0$ amounts to
\begin{equation}
	P_n'(1)+rP_n(1)=0 \,,
\end{equation}
hence
\begin{equation}
	r+\sum_{k=1}^n(k+r-rz_0^k)c_k=0 \,,
\end{equation}
and thus we obtain
\begin{equation}
	c_1 = \frac{-1}{{1+r-rz_0}}\left(r+\sum_{k=2}^n(k+r-rz_0^k)c_k\right) \,.
\end{equation}
Since the polynomial $P_n$ is not automatically guaranteed to be free of zeros inside the unit disk, one has to scan $r$ between $0$ and $\lfloor N/2\rfloor$, discard solutions with zeros inside the unit disk, and select from the remaining solutions the one with the best $\chi^2$. Overall, this implies that we still count $N-1$ free parameters for the conformal polynomial.

\subsection{Sum-rule constraint}
\label{sec:SumRuleConstraint}

Instead of using an explicit parametrization, the absence of zeros in the VFF can be  implemented by adding a term to the $\chi^2$ that penalizes zeros in the inelastic factor. In Ref.~\cite{Leutwyler:2002hm}, the absence of zeros in the VFF is used to write an unsubtracted dispersion relation for the function
\begin{equation}
	\label{eq:PsiFunction}
	\psi(s) = \frac{1}{(\sthr - s)^{3/2}} \log \frac{F_\pi^V(s)}{F_\pi^V(\sthr)} \, , \quad \sthr = 4M_\pi^2 \, .
\end{equation}
The normalization $F_\pi^V(0) = 1$ and the asymptotic behavior lead to the sum rules~\cite{Leutwyler:2002hm}
\begin{equation}
	\label{eq:SumRule32}
	\log F_\pi^V(\sthr) = \frac{1}{\pi}\int_{\sthr}^\infty\frac{ds}{s}\frac{\sthr^{3/2}}{(s-\sthr)^{3/2}}\log\left|\frac{F_\pi^V(s)}{F_\pi^V(\sthr)}\right|
\end{equation}
and
\begin{equation}
	\label{eq:SumRule12}
	\frac{\sqrt{\sthr}}{\pi}\int_{\sthr}^\infty\frac{ds}{(s-\sthr)^{3/2}}\log\left|\frac{F_\pi^V(s)}{F_\pi^V(\sthr)}\right|=0 \,.
\end{equation}
In our parametrization, only the inelastic factor $G_\inel(s)$ can develop zeros, and it satisfies similar analytic properties as the VFF, so we can use either of these relations with $G_\inel(s)$ on the inelastic branch cut only.  In the case of the second sum rule, we have
\begin{equation}
	\frac{\sqrt{s_\inel}}{\pi}\int_{s_\inel}^\infty\frac{ds}{(s-s_\inel)^{3/2}}\log\left|\frac{G_\inel(s)}{G_\inel(s_\inel)}\right|=0 \,.
\end{equation}
Using explicitly the parametrization of $G_\inel(s)$ as a conformal polynomial, a change of variable $e^{i\theta}=z(s)$ on the upper rim of the inelastic branch cut allows us to write this integral only in terms of the polynomial $P_N(z)$,
\begin{equation}
	I=\int_0^{\pi}\frac{d\theta}{\sin^2(\theta/2)}\log\left|\frac{P_N(e^{i\theta})}{P_N(1)}\right|=0 \,.
\end{equation}
We impose this constraint by adding to the $\chi^2$ function a penalty term
\begin{equation}
	\label{eq:Chi2PenaltyZeros}
	\chi^2_\text{zeros} = \frac{I^2}{\sigma_I^2} \, ,
\end{equation}
with $\sigma_I$ chosen small enough. We checked that this method leads to the same results as the explicit parametrization presented in Sect.~\ref{sec:ExplicitZeroFree}. However, it turns out that along certain paths in the parameter space, the penalty term~\eqref{eq:Chi2PenaltyZeros} discontinuously drops to zero. This makes it non-trivial to find the global minimum of the $\chi^2$ function, depending on the initial choice of parameters in the iterative fit routine. In practice, in order to arrive at stable fit results we make use of a combination of both the explicit zero-free parametrization and the sum-rule method. In the following, we refer with ``constrained fits'' to both of them equivalently.

\subsection{Inelasticity parameter}
\label{sec:InelasticityParameter}

In Refs.~\cite{Colangelo:2018mtw,Colangelo:2022prz,Stoffer:2023gba}, the second largest source of systematic uncertainties after the variation of the order $N$ of the conformal polynomial turned out to be the effect of the parameter $\iota_1$, which is used to describe the elasticity factor
\begin{equation}
    \eta_1(s)=\frac{s_a^3-\iota_1(s-4M_\pi^2)^{3/2}(s-s_\inel)^{3/2}}{s_a^3+\iota_1(s-4M_\pi^2)^{3/2}(s-s_\inel)^{3/2}},
\end{equation}
with $s_a=(1\,\text{GeV})^2$. A vanishing value of $\iota_1$ implies $\eta_1 = 1$ and due to the \EL{} bound~\eqref{eq:ELbound} a vanishing inelastic phase. In the Roy-equation analysis of Refs.~\cite{Ananthanarayan:2000ht,Caprini:2011ky}, the value $\iota_1 = 0.05(5)$ was used. However, as observed in Ref.~\cite{Colangelo:2018mtw} very small values of $\iota_1$ constrain the inelastic phase of the VFF too much and lead to a bad $\chi^2$ in the fits to the $e^+e^-$ data, hence the value $\iota_1 = 0.05(4)$ was chosen.

Here, we apply an improved treatment of the inelasticity parameter $\iota_1$. We revert to the original choice of Ref.~\cite{Ananthanarayan:2000ht,Caprini:2011ky} and use a normal distribution $\iota_1=0.05(5)$ as prior $p[\iota_1]$. We compute the posterior distribution with Bayes' rule
\begin{equation}
    p[\iota_1|\chi^2]=\frac{p[\chi^2|\iota_1]\,p[\iota_1]}{p[\chi^2]} \, ,
\end{equation}
where $p[\chi^2|\iota_1]$ is a $\chi^2$ distribution, and $p[\chi^2]=\int p[\chi^2|\iota_1]p[\iota_1]d\iota_1$. The posterior standard deviation is computed as
\begin{equation}
	\sigma_{\iota_1}^2=\int(\iota_1-\bar{\iota}_1)^2p[\iota_1|\chi^2] d\iota_1 \, .
\end{equation}
These integrals are evaluated by sampling several values of $\iota_1$ within the range $\iota_1\in [0.0,0.1]$ and assuming $p[\chi^2|\iota_1] = 0$ for the unphysical values $\iota_1 \le 0$. For $\iota_1>0.1$, we use a linear extrapolation for $p[\chi^2|\iota_1]$ determined with $\iota_1\in\{0.09,0.1\}$, where the \EL{} bound is no longer effective. The error propagation for all derived quantities is performed using the posterior distribution. This solves the initial problem of the large uncertainties caused by very bad fit qualities for low values of $\iota_1$, when using the prior distribution only.

%% file: sections/ModulusRepresentation.tex

\section{Hybrid phase-modulus representation}
\label{sec:HighEnergyContinuation}

In the complex plane of the conformal variable $z$, defined in Eq.~\eqref{eq:ConfZ}, the inelastic branch cut is mapped to the unit circle. The conformal expansion in terms of a polynomial has a convergence radius that is limited by the nearest singularities and it can be used to systematically approximate inelastic effects inside the unit disk. In addition, the \EL{} bound restricts the phase of the inelastic factor $G_\inel(s)$ in the region around $s\approx1\GeV^2$. However, excited vector mesons with masses above $1\GeV$ lead to resonance poles on the second Riemann sheet, which in the plane of the conformal variable $z$ lie only slightly outside the unit disk. Since a low-order conformal polynomial does not reproduce this resonance structure, it cannot be used on the inelastic branch cut well above $s\approx1\GeV^2$. Therefore, the parametrization~\eqref{eq:Gin} of inelastic effects is valid only up to $s \approx 1\GeV^2$.

In the following, we want to extend the description of the VFF beyond the region $s\approx1\GeV^2$. The treatment of inelastic effects in dispersive descriptions of the VFF is a classic subject~\cite{Pham:1975at,Mohapatra:1977ht} and has also been addressed in Refs.~\cite{Chanturia:2022rcz,Heuser:2024biq}. Recent high-statistics data are available from BaBar~\cite{BaBar:2009wpw,BaBar:2012bdw}, with the highest bin reaching $\sqrt{s} = 3\GeV$, as well as CMD-3~\cite{CMD-3:2023alj,CMD-3:2023rfe}, which is limited to $\sqrt{s}<1.2\GeV$. Since the data at higher energies constrain the inelastic factor $G_\inel(s)$, it is interesting to employ a representation with a larger range of validity in order to check the compatibility of the low-energy data with a continuation beyond $1\GeV^2$. In the case of compatible data, it can be used to constrain further the low-energy representation and to determine the impact of heavier resonances and inelasticities on the analytic continuation and the charge radius, $\< r_\pi^2 \>$, which depends on the form factor on the whole branch cut $s\in[\sthr;\infty)$. Detailed investigations along these lines have been performed in Refs.~\cite{Ananthanarayan:2011xt,Ananthanarayan:2012tt,Ananthanarayan:2013zua,Ananthanarayan:2016mns,Ananthanarayan:2017efc,Ananthanarayan:2018nyx,Ananthanarayan:2020vum}. Here, we follow a different approach, which allows us to take all the available data into account, at the price of introducing a parametrization dependence. We will show that this representation leads to results that are compatible with our low-energy representation.

The narrow isoscalar $\phi$ resonance can be treated in analogy to the $\omega$ resonance~\cite{Hanhart:2012wi}, see Eq.~\eqref{eq:G_omega}, and we define
\begin{equation}
	\label{eq:G_omega_phi}
	G_{\omega\phi}(s) = G_\omega(s) \times G_\phi(s) \approx G_\omega(s) + G_\phi(s) - 1 \, .
\end{equation}
The isovector resonances $\rho'$, $\rho''$ and $\rho'''$ are much broader and more difficult to describe. They are usually fit with a combination of Gounaris--Sakurai (GS) functions~\cite{Gounaris:1968mw}, $\text{GS}_{\rho^i}(s):=\text{GS}(s;M_{\rho^i},\Gamma_{\rho^i})$ (see, e.g., Refs.~\cite{BaBar:2012bdw, CMD-3:2023alj}), which model explicitly the resonance structures visible in the data (omitting the $\omega$ and $\phi$ resonances, which are described by the factor~\eqref{eq:G_omega_phi}),
\begin{equation}
	\label{eq:GS}
	\tilde{F}(s) = \frac{\text{GS}_\rho(s)+\epsilon_{\rho'}\text{GS}_{\rho'}(s)+\epsilon_{\rho''}\text{GS}_{\rho''}(s)+\epsilon_{\rho'''}\text{GS}_{\rho'''}(s)}{1+\epsilon_{\rho'}+\epsilon_{\rho''}+\epsilon_{\rho'''}} \, .
\end{equation}
Although the modulus of this function fits the data very well, it does not fulfill the constraints of analyticity and unitarity, in particular, one cannot rely on its phase. 
While the phase of the VFF below the inelastic threshold is given by the elastic $\pi\pi$ phase shift due to Watson's theorem~\cite{Watson:1954uc}, the inelastic phase is a priori not known and only constrained by the \EL{} bound.
However, it has been known for a long time (under the name of ``modulus representation") that one can reconstruct dispersively the whole VFF from its modulus only~\cite{PhysRev.172.1645}, under the asumption that it has no zeros or that the location of potential zeros is known~\cite{Geshkenbein:1998gu,Leutwyler:2002hm}. By writing a dispersion relation on $\psi(s)$ given in Eq.~\eqref{eq:PsiFunction},\footnote{Using $\psi(s)$ instead of $(\sthr-s)^{-1/2}\log[F_\pi^V(s)/F_\pi^V(\sthr)]$ requires an additional sum rule but enforces the correct $P$-wave threshold behavior.} one can derive
\begin{equation}
	F_\pi^V(s) = \mathcal{DR}\Bigl[ |F_\pi^V|;\sthr\Bigr](s) \, ,
\end{equation}
where
\begin{align}
	\label{eq:defDR}
	\mathcal{DR}\Bigl[|F|;s_a\Bigr](s) &:= |F(s_a)|^{1-\left(\frac{s_a-s}{s_a}\right)^{3/2}} \times\exp\left\{ -\frac{s(s_a-s)^{3/2}}{\pi}\int_{s_a}^\infty ds'\frac{\log|F(s')/F(s_a)|}{s'(s'-s_a)^{3/2}(s'-s)}\right\} \,.
\end{align}
Recently, this method was used in Ref.~\cite{RuizArriola:2024gwb} to reconstruct the phase of the VFF in the range $s\in[\sthr;(2.5\GeV)^2]$ from BaBar data, by fitting them with a GS representation, which was then inserted into the modulus representation~\eqref{eq:defDR}.
Even though this implements the analyticity constraints on the VFF itself, in contrast to the Omn\`es representation discussed in Sect.~\ref{sec:DispersiveRepresentation} it does not include explicitly the knowledge of the elastic phase from the Roy equations, nor the \EL{} bound and the $P$-wave behavior at the inelastic threshold. For this reason, we prefer to keep the phase representation for the Omn\`es factor $\Omega_1^1(s)$ in our description, and use the modulus representation to describe the inelastic effects only. This amounts to applying Eq.~\eqref{eq:defDR} to the function $G_\inel(s)=F_\pi^V(s)/[\Omega_1^1(s)G_{\omega\phi}(s)]$, with a cut starting at the inelastic threshold,
\begin{equation}
	\label{eq:Fhybrid}
	F_\pi^V(s) = \Omega_1^1(s)\times G_{\omega\phi}(s)\times \mathcal{DR}\Bigl[ | G_\inel |; s_\inel \Bigr](s) \, .
\end{equation}
A similar hybrid phase-modulus representation has been used in Refs.~\cite{Ananthanarayan:2011xt,Ananthanarayan:2012tt,Ananthanarayan:2013zua,Ananthanarayan:2016mns,Ananthanarayan:2017efc,Ananthanarayan:2018nyx,Ananthanarayan:2020vum}, in a model-independent approach that does not rely on a specific parametrization for $|G_\inel|$, but only uses an estimate for a weighted integral over the squared modulus above the inelastic threshold. Given sufficiently high-quality data on the modulus of the VFF above the inelastic threshold, one could try to integrate data directly. Here, in analogy to Ref.~\cite{RuizArriola:2024gwb} we use the modulus of the GS representation~\eqref{eq:GS} to interpolate the data and we approximate
\begin{equation}
	| G_\inel(s) | = \left| \frac{F_\pi^V(s)}{\Omega_1^1(s)G_{\omega\phi}(s)} \right| \approx | \tilde G(s) | \, , \quad \tilde G(s) := \frac{\tilde F(s)}{\Omega_1^1(s)} \, ,
\end{equation}
leading to
\begin{equation}
	\label{eq:FhybridGS}
	F_\pi^V(s) \approx \Omega_1^1(s)\times G_{\omega\phi}(s)\times \mathcal{DR}\Bigl[ |\tilde G|;s_\inel \Bigr](s) \, .
\end{equation}

It is interesting to rearrange the terms in Eq.~\eqref{eq:FhybridGS} in a simpler way, see App.~\ref{app:Hybrid},
\begin{equation}
	F_\pi^V(s) \approx \bar{\omega}_1^1(s)\times G_{\omega\phi}(s)\times\mathcal{DR}\Bigl[ |\tilde{F}|;s_\inel \Bigr](s) \, ,
\end{equation}
with $\bar{\omega}_1^1(s)$ defined as
\begin{align}
	\label{omegabar11}
	\log\bar{\omega}_1^1(s) &= \left[\left(\frac{s-\sthr}{s_\inel-\sthr}\right)^{3/2}-\left(\frac{\sthr(s-s_\inel)}{s_\inel(s_\inel-\sthr)}\right)^{3/2}\right]i\,\delta_1^1(s_\inel) \nn
		&\quad +\frac{s(s_\inel-s)^{3/2}}{\pi}\int_{\sthr}^{s_\inel}ds'\frac{\delta_1^1(s')-\left(\frac{s'-\sthr}{s_\inel-\sthr}\right)^{3/2}\delta_1^1(s_\inel)}{s'(s_\inel-s')^{3/2}(s'-s)} \, ,
\end{align}
as it separates the dependence on the elastic phase $\delta_1^1(s)$ and on the parameters for the modulus above $s_\inel$. Furthermore, one can see from Eq.~\eqref{omegabar11} that there is no dependence on the continuation of $\delta_1^1(s)$ above the inelastic threshold~\cite{Ananthanarayan:2020vum}.

For the reconstruction of $G_\inel(s)$, we use a dispersion relation for the function $\frac{\log[G_\inel(s)/G_\inel(s_\inel)]}{s(s-s_\inel)^{3/2}}$, which can generate terms of $\O(s^{3/2})$ and $\O(s^{1/2})$ in the asymptotic expansion of $\log G_\inel(s)$. Since the inelastic factor $G_\inel(s)$ should converge to a constant at $s\to\pm\infty$ (see Sect.~\ref{sec:DispersiveRepresentation}), the $s^{3/2}$ and $s^{1/2}$ terms have to vanish. This amounts to imposing two sum rules analogous to Eqs.~\eqref{eq:SumRule32} and~\eqref{eq:SumRule12}, which we write as
\begin{align}
	\label{eq:GinSumRules}
	\mathfrak{SR}_{3/2}\bigl[ |G_\inel| \bigr] &:= \frac{s_\inel^{3/2}}{\pi}\int_{s_\inel}^\infty ds'\frac{\log|G_\inel(s')/G_\inel(s_\inel)|}{s'(s'-s_\inel)^{3/2}}-\log|G_\inel(s_\inel)|=0 \, , \nn
	\mathfrak{SR}_{1/2}\bigl[ |G_\inel| \bigr] &:= \frac{s_\inel^{1/2}}{\pi}\int_{s_\inel}^\infty ds'\frac{\log|G_\inel(s')/G_\inel(s_\inel)|}{(s'-s_\inel)^{3/2}}=0.
\end{align}
Taking $|G_\inel(s)|=|\tilde{G}(s)|$ on the inelastic branch cut, these sum rules could be used as a constraint in the fit to data. However, we find that this has a significant impact on the goodness of fit, which might be a sign of a model bias introduced by imposing an exact GS functional form. Instead, we implement the sum rules by introducing an additional correction factor in the parametrization of the modulus. Since two degrees of freedom are needed in general to enforce the two sum rules, we correct $|G_\inel(s)|$ as
\begin{equation}
	|G_\inel(s)|=|\tilde{G}(s)|\times\exp\left(a+b\frac{s_\inel}{s}\right) \, .
\end{equation}
The correction factor does not affect the asymptotic leading power of $\log|G_\inel(s)|$, which remains a constant as for $\log|\tilde{G}(s)|$, but $a$ and $b$ can be fixed to cancel the diverging powers in $\arg G_\inel(s)$, leading to
\begin{equation}
	\label{eq:def_a_b}
	\left\{
		\begin{array}{l}
		a = \mathfrak{SR}_{3/2}\bigl[ |\tilde G| \bigr] - \frac{3}{2}\mathfrak{SR}_{1/2}\bigl[ |\tilde G| \bigr] \,, \\[0.2cm]
		b = \mathfrak{SR}_{1/2}\bigl[ |\tilde G| \bigr] \, ,
		\end{array}
	\right.
\end{equation}
where $\mathfrak{SR}_{3/2}$ and $\mathfrak{SR}_{1/2}$ are computed with $|\tilde{G}|$ only. One can rewrite the impact of this correction on the overall VFF as
\begin{equation}
	\label{eq:HybridRepresentation}
	F_\pi^V(s)=\bar{\omega}_1^1(s)\times G_{\omega\phi}(s)\times\mathcal{DR}\Bigl[ |\tilde{F}|;s_\inel\Bigr](s)\times G_{\infty}(s) \, ,
\end{equation}
with
\begin{equation}
	\label{eq:Gasymp}
	\log\,G_\infty(s)=\left(a+b\frac{s_\inel}{s}\right)\left[1-\left(\frac{s_\inel-s}{s_\inel}\right)^{3/2}\right]-\frac{3}{2}b\left(\frac{s_\inel-s}{s_\inel}\right)^{3/2} \, .
\end{equation}
One can easily check that this factor satisfies $G_\infty(0)=1$ and that it has the expected $P$-wave behavior at the inelastic threshold $s_\inel$. It is important to note that this additional factor does not add any free parameter nor does it contribute to the number of degrees of freedom, since $a$ and $b$ are completely fixed by the sum rules. However, it has a major impact on the stability of the results and the uncertainties, especially on the pion charge radius, see Sect.~\ref{sec:charge_radius}.

The factor $G_{\infty}(s)$ factorizes into a product of the contributions from $\tilde F$ and $1/\Omega_1^1$,
\begin{equation}
	G_{\infty}(s) = G_{\infty}^{|\tilde F|}(s) \times G_{\infty}^{1/|\Omega_1^1|}(s) \, .
\end{equation}
The different factors in $F_\pi^V(s)$ can be rearranged as
\begin{equation}
	\label{eq:HybridRepresentation2}
	F_\pi^V(s)= \left( \bar{\omega}_1^1(s) \; G_{\infty}^{1/|\Omega_1^1|}(s) \right) \times G_{\omega\phi}(s) \times \left( \mathcal{DR}\Bigl[ |\tilde{F}|;s_\inel\Bigr](s) \; G_{\infty}^{|\tilde F|}(s) \right) \, .
\end{equation}
The first big bracket only depends on the elastic phase $\delta_1^1(s)$ below $s_\inel$, the second big bracket only depends on the modulus of the VFF above the inelastic threshold. The correction factors drive the phase of the first bracket asymptotically to zero, whereas the phase of the second bracket reaches $\pi$ asymptotically. Due to the split, each of the two terms involves a slowly converging sum rule $\mathfrak{SR}_{1/2}[ 1 / | \Omega_1^1 | ]$ and $\mathfrak{SR}_{1/2}[ | \tilde F| ]$, respectively. In order to avoid numerical instabilities, ideally the two big brackets in Eq.~\eqref{eq:HybridRepresentation2} are computed directly by subtracting the sum-rule-violating contributions. E.g., we note that the Omn\`es contributions to the sum-rule terms can be written as
\begin{align}
	\mathfrak{SR}_{3/2}\bigl[ |\Omega_1^1| \bigr] &= \frac{\sthr^{3/2} - s_\inel^{3/2}}{(s_\inel - \sthr)^{3/2}} \delta_1^1(s_\inel) + \frac{s_\inel^{3/2}}{\pi} \int_{\sthr}^{s_\inel} ds \frac{\delta_1^1(s) - \left( \frac{s-\sthr}{s_\inel-\sthr} \right)^{3/2} \delta_1^1(s_\inel)}{s(s_\inel-s)^{3/2}}  \, , \nn
	\mathfrak{SR}_{1/2}\bigl[ |\Omega_1^1| \bigr] &= -\frac{3}{2} \frac{s_\inel^{1/2}}{(s_\inel - \sthr)^{1/2}} \delta_1^1(s_\inel) + \frac{s_\inel^{1/2}}{\pi} \int_{\sthr}^{s_\inel} ds \frac{\delta_1^1(s) - \left( \frac{s-\sthr}{s_\inel-\sthr} \right)^{3/2} \delta_1^1(s_\inel)}{(s_\inel-s)^{3/2}}  \, .
\end{align}

%% file: sections/Fits.tex

\section{\boldmath Fits to $e^+e^-$ data}
\label{sec:Fits}

In the following, we perform fits of the different dispersive representations of the VFF to the high-statistics $e^+e^-\to\pi^+\pi^-$ data sets from SND~\cite{Achasov:2005rg,Achasov:2006vp,SND:2020nwa}, CMD-2~\cite{CMD-2:2001ski,CMD-2:2003gqi,Aulchenko:2006dxz,CMD-2:2006gxt}, BaBar~\cite{BaBar:2009wpw,BaBar:2012bdw}, KLOE~\cite{KLOE:2008fmq,KLOE:2010qei,KLOE:2012anl,KLOE-2:2017fda}, BESIII~\cite{BESIII:2015equ}, and CMD-3~\cite{CMD-3:2023alj,CMD-3:2023rfe}. As in previous work~\cite{Colangelo:2018mtw,Colangelo:2022prz,Stoffer:2023gba}, we exclude two points out of 195 from the KLOE data set (denoted by KLOE$''$), which were identified in Ref.~\cite{Colangelo:2018mtw} to barely affect the fit results but to give a huge contribution to the $\chi^2$. We also include the space-like data from NA7~\cite{NA7:1986vav} in our combined fits.

To gauge the impact of the absence of zeros on the HVP contribution to $a_\mu$ and the pion charge radius, we implement three different setups.
\begin{itemize}
	\item The ``unconstrained'' low-energy fits refer to a similar setup as Refs.~\cite{Colangelo:2018mtw,Colangelo:2022prz,Stoffer:2023gba}, i.e., the low-energy description presented in Sect.~\ref{sec:DispersiveRepresentation} with a conformal polynomial constrained only by the \EL{} bound. They are used as a baseline in this analysis. Compared to previous work, we include the improved treatment of the inelasticity parameter $\iota_1$ as described in Sect.~\ref{sec:InelasticityParameter}.
	\item The ``constrained'' low-energy fits exclude the presence of complex zeros by constraining the conformal polynomial in addition to the \EL{} bound with one of the methods discussed in Sects.~\ref{sec:ExplicitZeroFree} and~\ref{sec:SumRuleConstraint}.
	\item The ``hybrid'' fits refer to the description presented in Sect.~\ref{sec:HighEnergyContinuation} and include the continuation to energies above $1\GeV$. In this setup, the \EL{} bound turns out to be automatically satisfied.
\end{itemize}

As in Ref.~\cite{Colangelo:2018mtw}, the fit parameters include in all three setups the values of the elastic phase at the two points $s_0$ and $s_1$, which parametrize the Roy solutions, the $\omega$ mass, real and imaginary parts of the $\rho$--$\omega$ mixing parameter $\epsilon_\omega$, as well as an energy-rescaling factor for each experiment, constrained by the experimental calibration uncertainty, see Ref.~\cite{Colangelo:2018mtw}. In the two low-energy setups, we fit in addition the coefficients of the conformal polynomial, whereas in the hybrid setup we fit the GS parameters (masses and widths of $\rho$, $\rho'$, $\rho''$, $\rho'''$, as well as real and imaginary parts of the couplings of the three excited resonances). The two GS parameters $M_\rho$ and $\Gamma_\rho$ enter the hybrid representation only indirectly through the tail of the $\rho$ resonance in the modulus above $s_\inel$ and hence they are poorly determined. To stabilize the fit, we include for these two parameters a prior in the $\chi^2$ obtained from a pure GS fit to BaBar data. Due to the presence of the correction factor~\eqref{eq:Gasymp}, the remaining GS parameters in the hybrid representation do not need to agree with the ones obtained in a pure GS fit. It rather turns out that the best fit prefers parameter values that lead to a rather large correction factor and hence a significant deviation from the GS functional form.

For the width of the $\omega$ meson, as well as the mass and width of the $\phi$ meson, we take as input the vacuum-polarization-subtracted values from the three-pion channel~\cite{Hoferichter:2023bjm},
\begin{equation}
	\Gamma_\omega = 8.71(3)\MeV \, , \quad M_\phi = 1019.21(2)\MeV \, , \quad  \Gamma_\phi = 4.249(13)\,\text{MeV} \, .
\end{equation}
Our analysis of systematic errors in the dispersive description follows previous work, see Ref.~\cite{Colangelo:2018mtw} for details. It involves the variation of all input parameters, including all the parameters that describe the Roy-equation solution for $\delta_1^1$. In the low-energy fits, further systematic uncertainties are related to the continuation of $\delta_1^1$ above $s_\inel$~\cite{Colangelo:2018mtw}. Finally, the dominant systematic uncertainty in the unconstrained fits is due to the variation of the order of the conformal polynomial $N$. In the unconstrained fits, we vary the number of free parameters in the conformal polynomial in the range $N-1=2\ldots5$. For $N-1>5$, clear signs of overfitting can be observed~\cite{Colangelo:2018mtw}, in particular in most cases no further improvement of the $p$-value can be obtained. In contrast to Refs.~\cite{Colangelo:2018mtw,Colangelo:2022prz,Stoffer:2023gba}, we exclude the case $N-1=1$. While for most experiments this has no impact on the uncertainty estimate, in the case of CMD-3 it leads to a dramatic reduction of the systematic uncertainty. The $p$-value demonstrates that the inclusion of $N-1=1$ in the preliminary fits to CMD-3 data of Ref.~\cite{Stoffer:2023gba} overestimates the systematic uncertainties: it jumps from $2.6\%$ for $N-1=1$ to $12.7\%$ for $N-1=2$ and stays around $20\%$ for larger values of $N-1$. Therefore, here we use the range $N-1=2\ldots5$, both for the unconstrained and constrained low-energy fits. As in previous work, we take $N-1=4$ for the central results of the unconstrained fits. In the case of the constrained fits, the $p$-value reaches a maximum at $N-1=3$ and remains roughly constant or slightly decreases for higher values of $N-1$, hence we use $N-1=3$ as central value in the constrained fits.

\begin{table}[t]
    \centering
    \input{tables/detailed_single_exp}
    \caption{Comparison of the unconstrained fits (without brackets), constrained fits (with square brackets), and fits using the hybrid representation~\eqref{eq:HybridRepresentation} (with parentheses). The first error is the fit uncertainty, inflated by $\sqrt{\chi^2/\text{dof}}$, the second error is the combination of all systematic uncertainties. The third error in the combined fits (which include NA7 and all $e^+e^-$ data sets apart from SND20 and CMD-3) takes into account the BaBar--KLOE tension with the WP prescription~\cite{Aoyama:2020ynm}.}
    \label{tab:detailed_single_exp}
\end{table}

The results for the unconstrained and constrained low-energy fits and the hybrid fits to single experiments are shown in Table~\ref{tab:detailed_single_exp}, in terms of the relevant low-energy parameters.
In the case of the hybrid fit to BaBar, we use the entire data set up to $\sqrt{s} = 3\GeV$, including the full covariance matrices. The hybrid fits to the other experiments are performed by fitting these low-energy data sets in combination with the BaBar data above $\sqrt{s} = 1.4\GeV$. This should lead to fit results that are largely dominated by the single low-energy experiment and not by BaBar. However, a comparison of the hybrid fit with the low-energy setups shows that the BaBar input at higher energies still has some impact and slightly pulls the fit results towards the BaBar values. While most experiments have data only below $1\GeV$, CMD-3 reaches $1.2\GeV$ and hence covers the $\phi$ resonance region. In the fits to the other low-energy data sets, we do not include the $\phi$ resonance, which has a negligible effect on $a_\mu$ or the pion charge radius.

The systematic uncertainties in the unconstrained fits are slightly reduced compared to Ref.~\cite{Stoffer:2023gba}, due to the improved treatment of $\iota_1$. As explained above, in the case of CMD-3 one can observe a more drastic reduction of the systematic uncertainty, which arises because the $N-1=1$ fit with low $p$-value is excluded. Comparing the constrained with the unconstrained fits in Table~\ref{tab:detailed_single_exp}, we find compatible results within uncertainties. The fit uncertainties as well as the systematic uncertainties on the $\omega$-resonance parameters remain largely unaffected by the constraint on the conformal polynomial. However, the systematic uncertainty on $a_\mu$ is significantly reduced in the constrained low-energy fits. The $p$-value of the constrained fits is only slightly lower than that of the unconstrained fits and remains very good in almost all cases. (The only exception are the fits to SND20~\cite{SND:2020nwa}, which lead to a poor $p$-value already in the unconstrained fits~\cite{Colangelo:2022prz}.) We interpret these observations as a clear sign that the data do not prefer zeros in the VFF, and we take full advantage of the much more stable fits when imposing the absence of zeros.

The hybrid fit to CMD-3 and BaBar data above $1.4\GeV$ has a $p$-value of $0.9\%$, much lower than the unconstrained and constrained low-energy fits to CMD-3, which reach almost $20\%$. This indicates that the CMD-3 data is in some tension even with BaBar data above $1.4\GeV$. If CMD-3 data only below $1\GeV$ are included, the $p$-value reaches $31\%$, hence the dispersive constraints do not indicate any tension between low-energy CMD-3 data and BaBar data above $1.4\GeV$. We conclude that the CMD-3 data above $1\GeV$ generate the observed tension with BaBar data above $1.4\GeV$.

\section{Results}
\label{sec:Results}

\subsection[Low-energy contribution to $a_\mu$]{\boldmath Low-energy contribution to $a_\mu$}

\begin{table}[t]
    \centering
    \input{tables/a_mu_vs_N}
    \caption{Comparison of the values of $a_\mu^{\pi\pi}|_{\leq1\text{GeV}}$ (in units of $10^{-10}$) in the unconstrained fits (without brackets) and constrained fits (with brackets) to single experiments, for different values of $N$. For the fixed-$N$ results, we use $\iota_1=0.05$ and the errors are the fit uncertainties, inflated by $\sqrt{\chi^2/\text{dof}}$. The central values are reproduced from Table~\ref{tab:detailed_single_exp}.}
    \label{tab:a_mu}
\end{table}

\begin{figure}[t]
    \centering
    \includegraphics[scale=0.5]{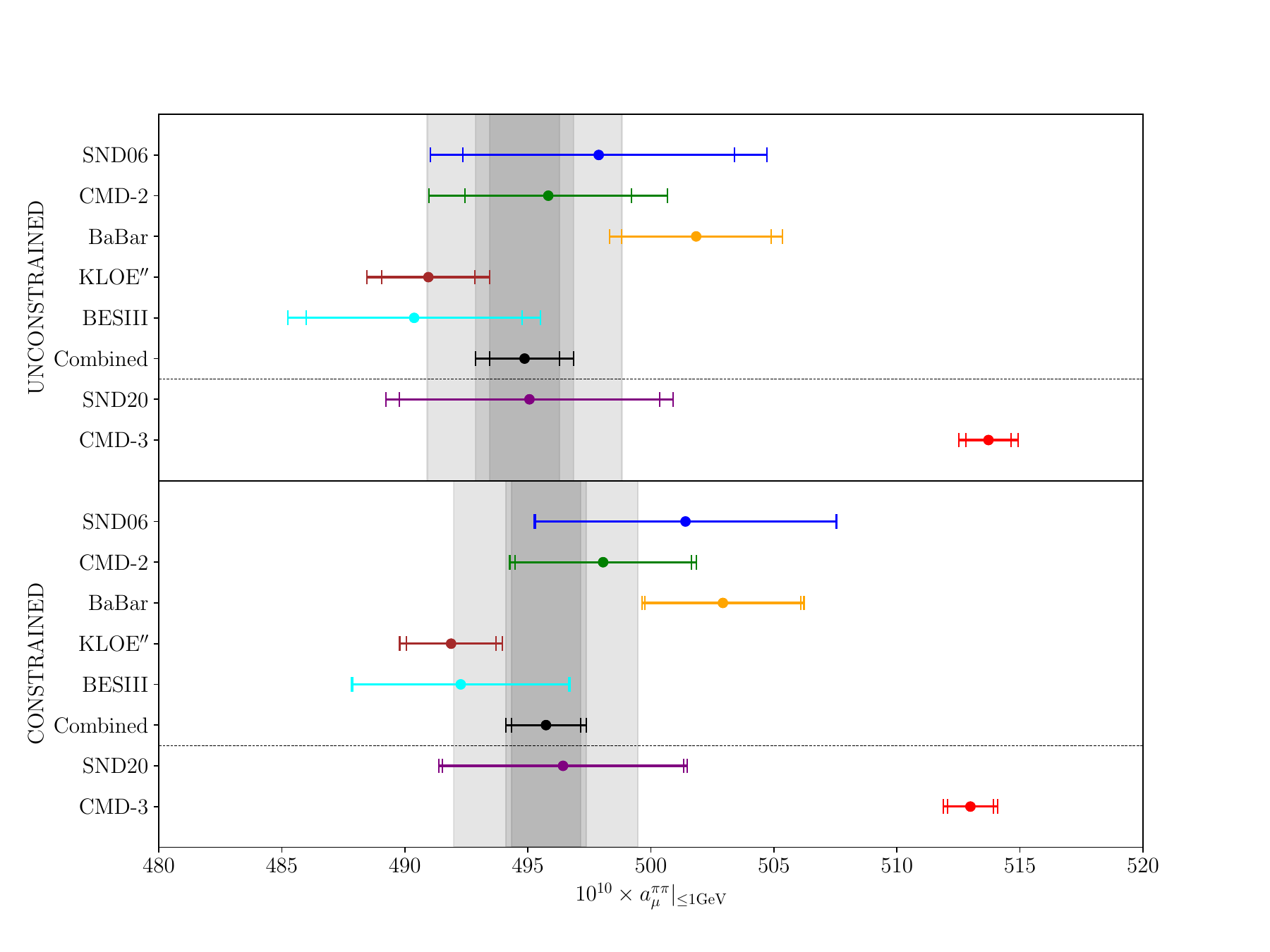}
    \caption{Values of $a_\mu^{\pi\pi}|_{\leq1\text{GeV}}$ from the unconstrained (above) and constrained (below) low-energy fits to single experiments. The smaller error bars show the fit uncertainties while the larger error bars show the full uncertainties including both fit uncertainties and systematic uncertainties. The gray bands correspond to the combined fit to all $e^+e^-$ data sets (apart from SND20 and CMD-3) and the NA7 data set, with the largest band including the additional systematic effect due to the BaBar–KLOE tension.}
    \label{fig:a_mu_1GeV_default_nozero}
\end{figure}

The dependence of $a_\mu^{\pi\pi}|_{\leq1\text{GeV}}$, the two-pion contribution to $a_\mu$ below $1\GeV$, on the number of free parameters in the conformal polynomial is shown in Table~\ref{tab:a_mu} for both constrained and unconstrained low-energy fits in the range $N-1=1\ldots7$. It is clearly visible that the constrained results are much more stable when $N$ varies, and especially in the range of interest $N-1=2\ldots5$. Figure \ref{fig:a_mu_1GeV_default_nozero} shows a comparison of the values of $a_\mu^{\pi\pi}|_{\leq1\text{GeV}}$ in the constrained and unconstrained low-energy fits. The reduction of the systematic uncertainties in the constrained fits is clearly visible. The fit uncertainties remain almost the same and dominate in the constrained fits to all data sets.

\begin{figure}[t]
    \centering
    \includegraphics[scale=0.64]{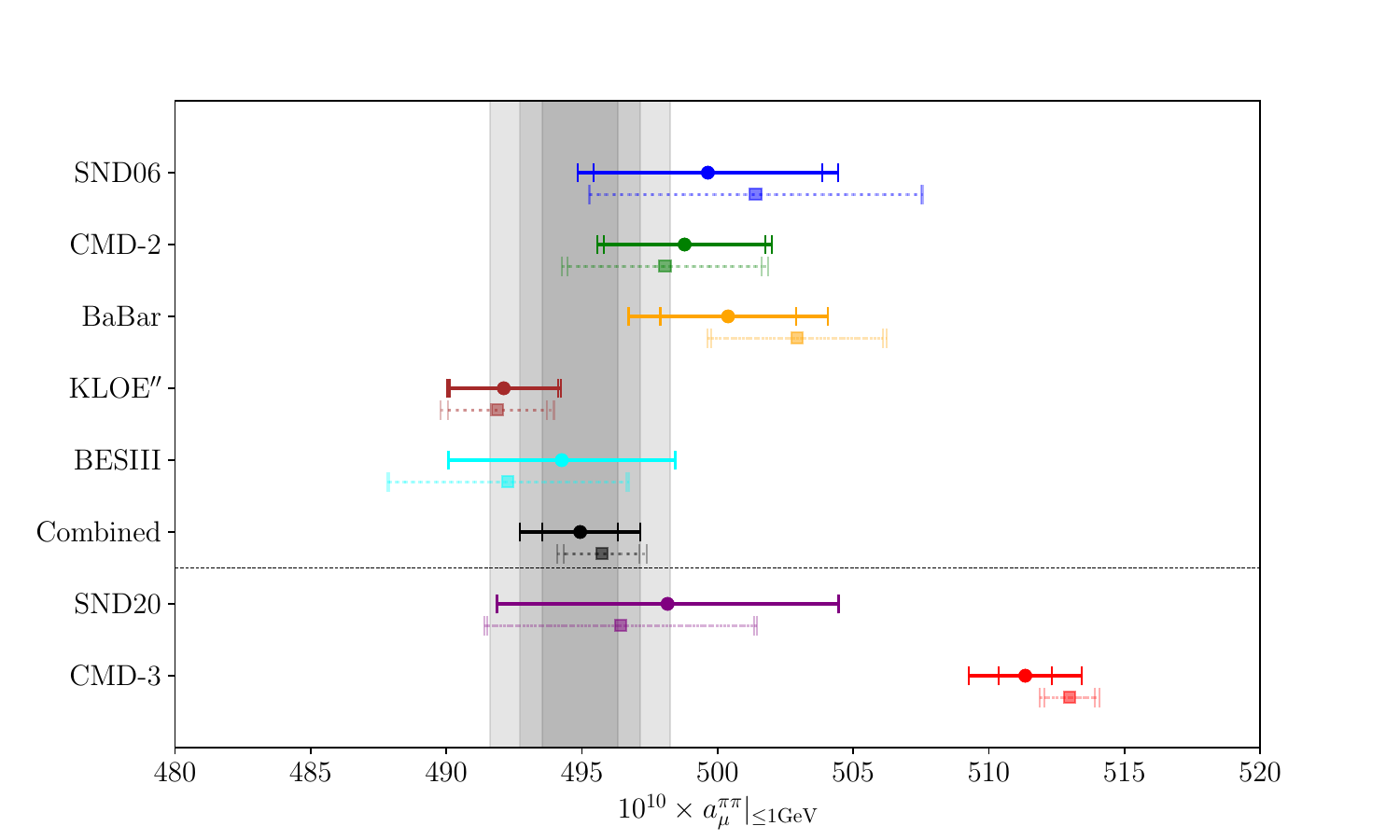}
    \caption{Values of $a_\mu^{\pi\pi}|_{\leq1\text{GeV}}$ from the hybrid fits to single experiments and BaBar data above $1.4\GeV$. The results of the low-energy constrained fits are shown for comparison as transparent points with dotted error bars. See also Fig.~\ref{fig:a_mu_1GeV_default_nozero} caption.}
    \label{fig:a_mu_1GeV_highres}
\end{figure}

The results for the fits of the hybrid representation are presented in Fig.~\ref{fig:a_mu_1GeV_highres}. They are fully consistent within uncertainties with the low-energy fits. Therefore, we will keep the low-energy fits as the central results for $a_\mu^{\pi\pi}|_{\leq1\text{GeV}}$, as they involve only data from single experiments. In addition to the fits to single experiments, we perform a fit to a combination of all $e^+e^-$ experiments apart from SND20 and CMD-3, as well as the space-like data from NA7. As in Refs.~\cite{Colangelo:2022prz,Stoffer:2023gba}, we exclude SND20 from the combination due to the poor $p$-value of the fit to this single data set. For this combined fit, we find
\begin{equation}
	a_\mu^{\pi\pi}|_{\leq1\text{GeV}}^\text{comb} = 495.7(1.4)_{\text{stat}}(0.8)_{\text{syst}}(3.3)_{\text{BK}}\times10^{-10} = 495.7(3.7)\times10^{-10} \, ,
\end{equation}
where the first error denotes the fit uncertainty, the second one the systematic uncertainty, and the third the additional systematic uncertainty due to the BaBar--KLOE tension, determined according to the WP prescription~\cite{Aoyama:2020ynm}. In contrast, the constrained fit to CMD-3 leads to
\begin{equation}
	a_\mu^{\pi\pi}|_{\leq1\text{GeV}}^\text{CMD-3} = 513.0(0.9)_{\text{stat}}(0.8)_{\text{syst}}\times10^{-10} = 513.0(1.2)\times10^{-10} \, .
\end{equation}
The discrepancies of all the individual experiments with CMD-3 are shown in Table~\ref{tab:discrepancies_amu}. Is is important to note that the dispersive representation enhances the discrepancies in $a_\mu$: since the VFF has to fulfill the fundamental constraints of analyticity and unitarity, the discrepancies in the energy intervals measured by the experiments imply discrepancies also outside these intervals, as the cross section cannot exhibit sudden unphysical jumps. In our framework, these constraints of analyticity and unitarity are implemented in the dispersive fit function. Hence, it is no surprise that we find much larger discrepancies compared to a strategy where the cross-section data of a particular experiment are supplemented outside the measured interval by a global average~\cite{Davier:2023fpl}. We note in particular that the tension between KLOE and CMD-3 reaches the level of $9\sigma$. The general conclusions hold irrespective of the assumption about the absence of zeros in the VFF---in most cases, imposing the absence of zeros increases the tensions further.

\begin{table}[t]
    \centering
    \input{tables/discrepancies_amu}
    \caption{Significance of the discrepancies between fits to CMD-3 and the other experiments, taking into account the correlations due to the systematics in the dispersive representation, as well as the $\chi^2$ inflation of the fit errors. For the combined fit (defined as before), the discrepancies in square brackets exclude the systematic effect due to the BaBar–KLOE tension. The Euclidean windows are evaluated with the constrained fits up to $1\GeV$.}
    \label{tab:discrepancies_amu}
\end{table}

\begin{figure}[t]
    \centering
    \includegraphics[scale=0.64]{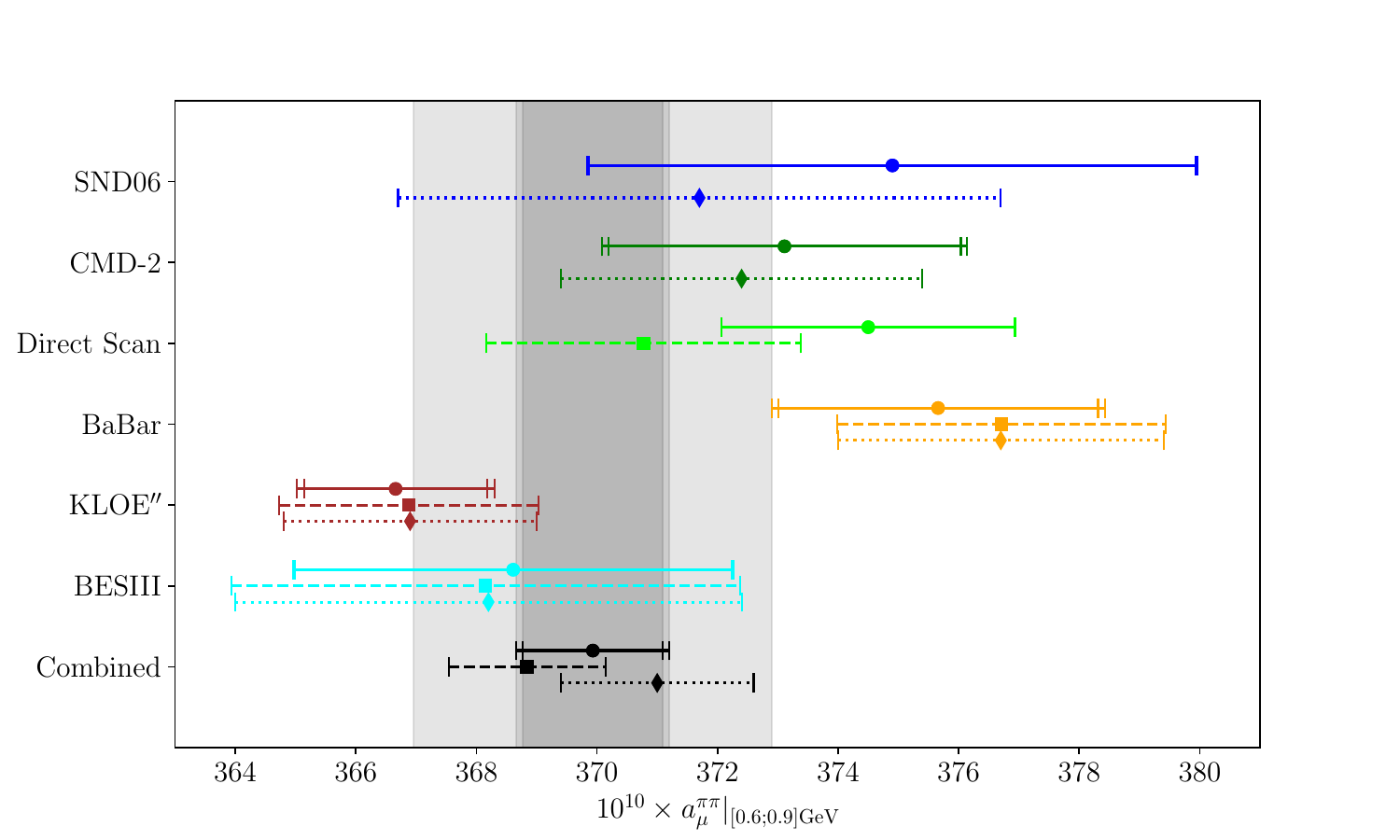}
    \caption{Values of $a_\mu^{\pi\pi}|_{[0.6;0.9]\GeV}$ from the constrained low-energy fits to single experiments (circles with plain error bars), compared to the evaluations of DHMZ~\cite{Davier:2019can} (diamonds with dotted error bars) and KNT~\cite{Keshavarzi:2019abf} (squares with dashed error bars). The direct-scan result in our fits includes SND06, CMD-2, and the NA7 data sets. See also Fig.~\ref{fig:a_mu_1GeV_default_nozero} caption.}
    \label{fig:a_mu_comparison_DHMYZ_KNT}
\end{figure}

In the first three columns of Table~\ref{tab:a_mu_ranges} and in Fig.~\ref{fig:a_mu_comparison_DHMYZ_KNT}, we show a comparison with some of the existing results from Refs.~\cite{Aoyama:2020ynm,Ananthanarayan:2018nyx,Colangelo:2022vok,Davier:2019can,Keshavarzi:2019abf}. In general, good agreement within uncertainties is observed. As in previous analyses~\cite{Colangelo:2018mtw,Colangelo:2022prz,Stoffer:2023gba}, our result is in good agreement with Ref.~\cite{Ananthanarayan:2018nyx} in the very low-energy region. Implementing the constraints of analyticity and unitarity in a rigorous way and imposing the absence of zeros, our approach leads to similar or smaller uncertainties than a direct integration of the data. Without imposing the absence of zeros, this effect was limited mainly to the low-energy region~\cite{Colangelo:2018mtw} due to the dominance of the systematic uncertainty from the variation of $N$.

\begin{table}[t]
    \centering
    \input{tables/a_mu_ranges}
    \caption{Values of $a_\mu^{\pi\pi}$ for different integration ranges. The first three columns are obtained from the constrained low-energy fits, whereas the last four columns are based on the hybrid representation. The combined fit and uncertainties of our results are defined as in Table~\ref{tab:detailed_single_exp}. Results from Refs.~\cite{Aoyama:2020ynm,Ananthanarayan:2018nyx,Colangelo:2022vok,Davier:2019can,Keshavarzi:2019abf} are shown for comparison.}
    \label{tab:a_mu_ranges}
\end{table}

\subsection{\boldmath The full two-pion channel}

\begin{figure}[t]
    \centering
    \includegraphics[scale=0.6]{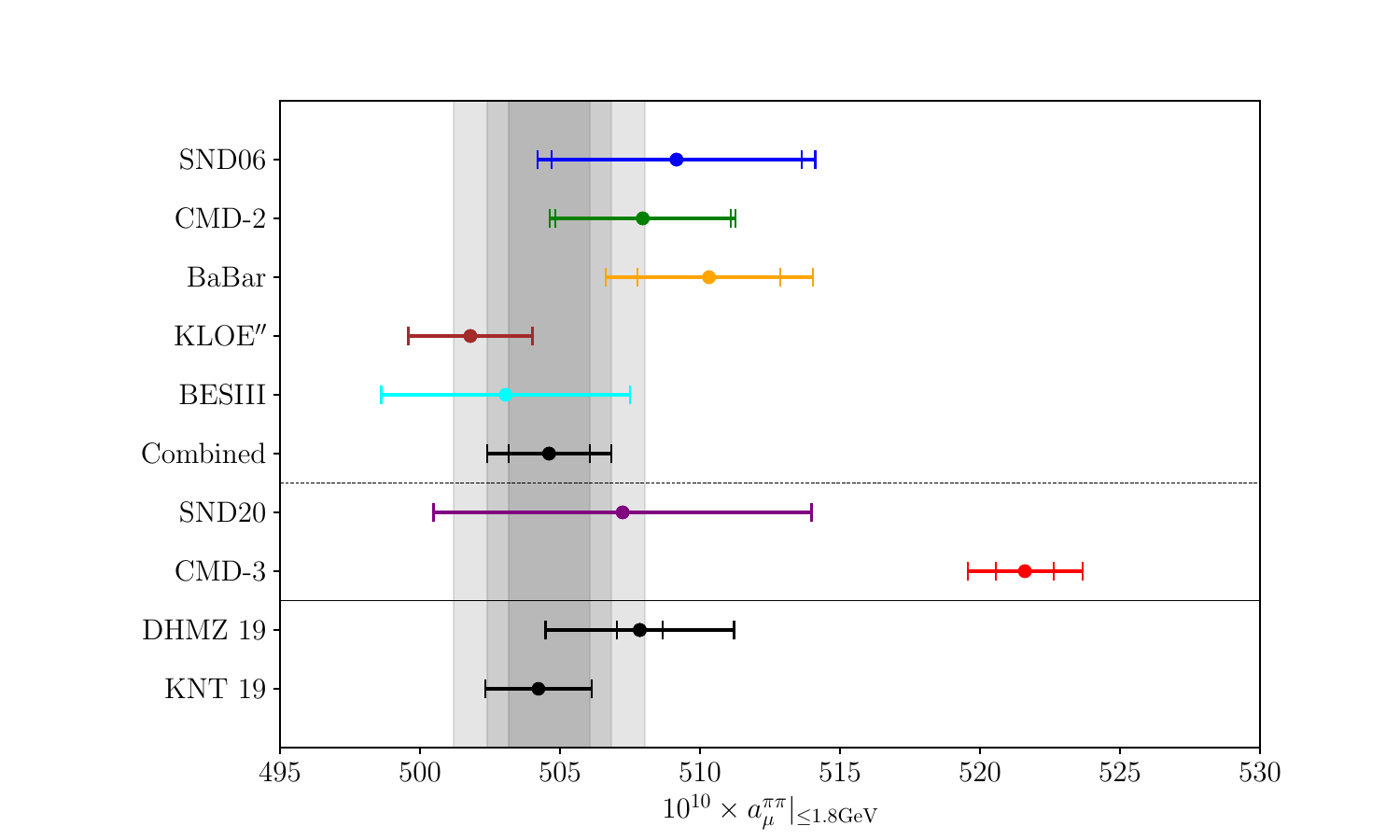}
    \caption{Values of $a_\mu^{\pi\pi}|_{\le1.8\GeV}$ from the hybrid fits to single experiments and BaBar data above $1.4\GeV$. Error bars and combination are defined as in Fig.~\ref{fig:a_mu_1GeV_default_nozero}. Results from DHMZ 19~\cite{Davier:2019can} and KNT 19~\cite{Keshavarzi:2019abf} are shown for comparison.}
    \label{fig:a_mu_full_highres}
\end{figure}

The hybrid representation of the VFF presented in Sect.~\ref{sec:HighEnergyContinuation} is constrained by data up to $3\GeV$. The contribution from energies above $3\GeV$ is completely negligible, hence this representation allows us to determine the complete two-pion contribution $a_\mu^{\pi\pi}$. Our final estimate
combining all $e^+e^-$ experiments (apart from SND20 and CMD-3) as well as spacelike data from NA7 reads
\begin{equation}
	a_\mu^{\pi\pi}|^\text{comb} = 504.7(1.2)_{\text{stat}}(1.6)_{\text{syst}}(2.9)_{\text{BK}}\times10^{-10} = 504.7(3.5)\times10^{-10} \, ,
\end{equation}
in good agreement with evaluations by DHMZ~\cite{Davier:2019can} and KNT~\cite{Keshavarzi:2019abf}.
The hybrid fit to CMD-3 and BaBar data above $1.4\GeV$ leads to
\begin{equation}
	a_\mu^{\pi\pi}|^\text{CMD-3} = 521.8(0.9)_{\text{stat}}(1.5)_{\text{syst}}\times10^{-10} = 521.8(1.7)\times10^{-10} \, .
\end{equation}
In compilations of the full HVP contribution~\cite{Davier:2019can,Keshavarzi:2019abf}, the exclusive two-pion channel is used only below a certain energy threshold and perturbative QCD or inclusive data are used above. To allow direct comparison, in Table~\ref{tab:a_mu_ranges} we provide our results for energy ranges that are cut at $1.8\GeV$, $1.937\GeV$, and $3\GeV$. Our results below $1.8\GeV$ are plotted in Fig.~\ref{fig:a_mu_full_highres}, in comparison with Refs.~\cite{Davier:2019can,Keshavarzi:2019abf}.

\subsection{Euclidean windows}
\label{sec:EuclideanWindows}

\begin{table}[t]
    \centering
    \input{tables/a_mu_windows}
    \caption{Two-pion contributions to Euclidean windows. The upper half of the table shows the window contributions integrated up to $s=1\GeV^2$, obtained from the constrained low-energy fits. The lower half of the table shows the complete two-pion contribution, using the hybrid representation. The sum of SD, int, and LD windows gives the total, whereas the VLD window is the fraction of the LD window above $t_\text{cut} = 2.8\fm$.}
    \label{tab:a_mu_windows}
\end{table}

In the context of lattice-QCD computations, partial HVP contributions that correspond to windows in Euclidean time~\cite{RBC:2018dos,Lehner:2020crt} have proven invaluable, both for a more in-depth comparison of different calculations, but also for a splitting into regions that pose different technical challenges. Defined by smooth step functions, three Euclidean windows have become standard, which split the HVP contribution at the two points
\begin{equation}
	t_0 = 0.4\fm \, , \quad t_1=1.0\fm
\end{equation}
into a short-distance (SD) window below $t_0$, an intermediate window from $t_0$ to $t_1$, and a long-distance (LD) window above $t_1$, with smoothing distance $\Delta=0.14\fm$. The step functions in Euclidean time can be transformed into weight functions in the dispersive integral in $s$, which allows one to evaluate the window quantities dispersively with data input~\cite{Colangelo:2022vok}. Several other windows are in use. Here, in addition to the three standard windows we consider a very-long-distance (VLD) window defined by $t>t_\text{cut} = 2.8\fm$. In this region, Ref.~\cite{Boccaletti:2024guq} uses data-driven input to complement the lattice-QCD computation below $t_\text{cut}$ by
\begin{equation}
	a_\mu^\text{VLD}|_\text{\cite{Boccaletti:2024guq}} = 27.59(26) \times 10^{-10} \, .
\end{equation}

Our results for the two-pion contribution to the window quantities are shown in Table~\ref{tab:a_mu_windows}, in comparison with the results from Ref.~\cite{Colangelo:2022vok}, which are in good agreement. In addition to the contribution below $1\GeV$, which we evaluate based on the constrained low-energy fits, we also show the complete two-pion contribution obtained with the hybrid representation. In Table~\ref{tab:discrepancies_amu}, we also show the discrepancies in the window quantities between the fit to CMD-3 and the other experiments. In general, the discrepancies in the window quantities are comparable to the ones in the total two-pion contribution.

\begin{figure}[t]
    \centering
    \includegraphics[scale=0.5]{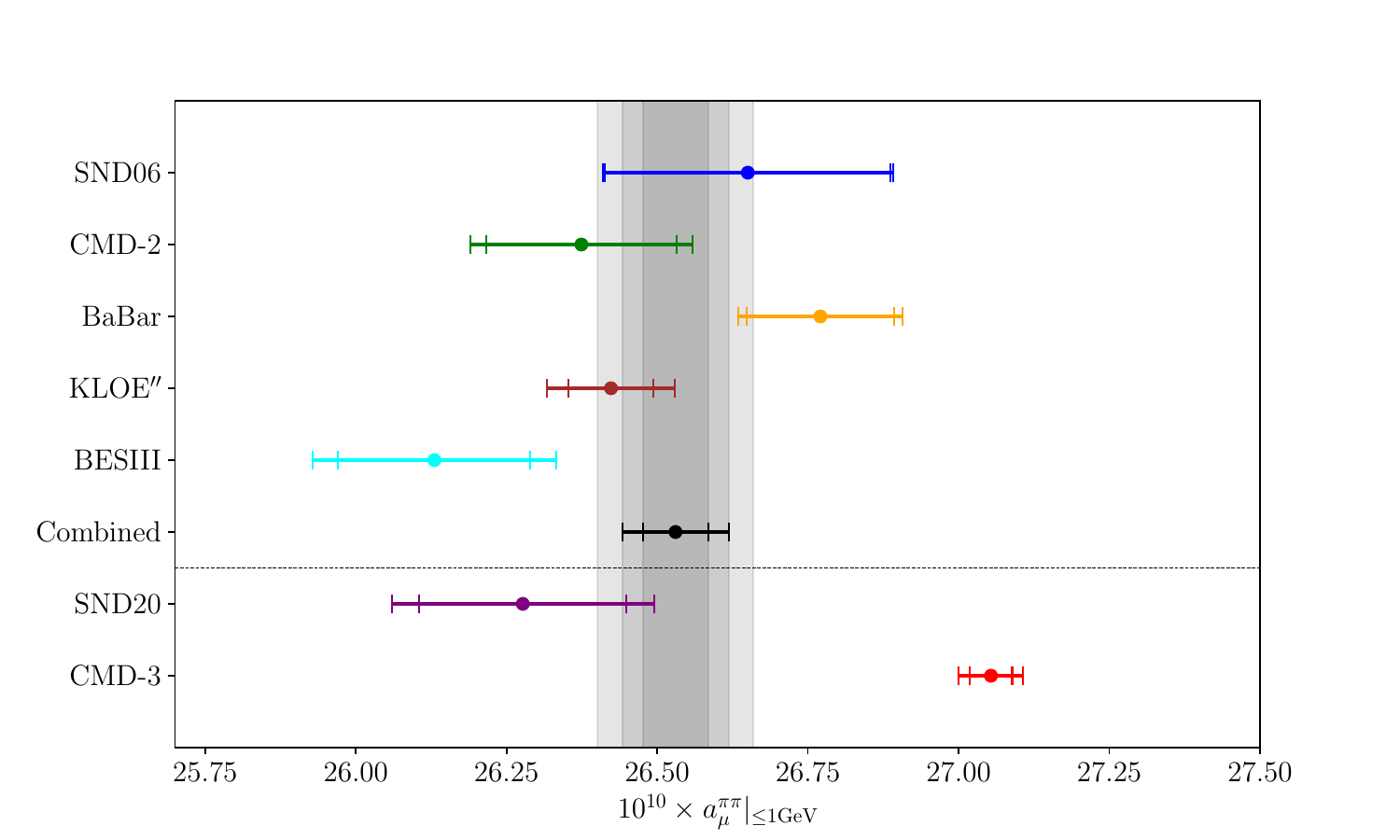}
    \caption{Two-pion contribution to the VLD window in Euclidean time ($t>2.8\fm$), obtained from the constrained low-energy fits.}
    \label{fig:VLDwindow}
\end{figure}

We observe that the constraints of unitarity and analyticity imply that discrepancies between the different $e^+e^-$ experiments persist even in the VLD window above $2.8\fm$. As shown in Table~\ref{tab:discrepancies_amu}, the discrepancy between KLOE and CMD-3 in the VLD window still lies above $5\sigma$. In all our fits, this window is indeed correlated a posteriori by more than $80\%$ with the intermediate window, where some of the largest discrepancies between the data sets are observed. As shown in Fig.~\ref{fig:VLDwindow}, the largest spread between the different results is almost one unit in $10^{-10}$ (and still $0.63\times10^{-10}$ between KLOE and CMD-3). Although this is much smaller than the total uncertainty of the recent BMWc update~\cite{Boccaletti:2024guq}, it exceeds the uncertainty of $0.26\times10^{-10}$ assigned to their data-driven determination of the VLD window contribution.

\subsection{Pion charge radius}
\label{sec:charge_radius}

A dispersive representation of the VFF allows a direct evaluation of the pion charge radius,
\begin{equation}
	\< r_\pi^2\> = 6\left.\frac{d F_\pi^V(s)}{d s}\right|_{s=0} \, ,
\end{equation}
since the dispersion relation provides the analytic continuation from the time-like region to $s=0$ and also into the space-like region. In practice, it is best to evaluate the charge radius via sum rules, see App.~\ref{app:Radius}. Given the high precision of $e^+e^-$ cross-section data, dispersive determinations of the charge radius from the time-like region~\cite{Ananthanarayan:2017efc,Colangelo:2018mtw} in general are more precise than extrapolations from the space-like region~\cite{Simula:2023ujs}. In Ref.~\cite{Colangelo:2020lcg}, the potential of the charge radius was analyzed as a tool to distinguish between $e^+e^-$ cross sections existing at that time and hypothetical modified cross sections that would reproduce lattice-QCD results for the HVP contribution to $a_\mu$ but would still be in line with the constraints of unitarity and analyticity. In the meantime, with the CMD-3 data set we have a concrete realization of such a different cross section. It indeed corresponds to one of the possible scenarios identified in Ref.~\cite{Colangelo:2020lcg}, a rather uniform shift in the cross section that leaves the low-energy parameters, i.e., the values of the elastic phase at $s_0$ and $s_1$, largely unaffected.

Table~\ref{tab:r_pi2_vs_N} shows the dependence of $\< r_\pi^2\>$ on the number of free parameters $N-1$ in the conformal polynomial for both constrained and unconstrained low-energy fits to single experiments. The range $N-1=3\ldots7$ is particularly interesting: in the unconstrained fits, it is the range where the results vary the most. This instability is connected with the appearance of zeros in the inelastic factor at $N-1=4$. In contrast, by avoiding zeros in the form factor the constrained fits are much more stable in the entire range.

Due to the instabilities in the unconstrained fits for higher $N$, in Ref.~\cite{Colangelo:2018mtw} the fit with $N-1=1$ was selected as central result for the charge radius. Here, we choose instead the $N-1=3$ fit results as preferred central values: they agree between unconstrained and constrained fits and lead to the best $p$-value in the constrained fits. As before, the variation over the range $N-1=2\ldots5$ is taken as a systematic uncertainty. Since this variation is very asymmetric, assigning a conservative symmetric error from the largest variation leads to somewhat different conclusions than a treatment with asymmetric errors.

Finally, the hybrid phase-modulus representation of the VFF provides similar results as the constrained low-energy fits, often with even smaller error bars. This is explained by a better control over the inelastic factor with the modulus dispersion relation, whereas in the low-energy fits, the variation of $N$ introduces the dominant uncertainty. Furthermore, the hybrid fits do not have a strong dependence on $\iota_1$ since the \EL{} bound is automatically fulfilled in this representation. In addition, the constraints provided by the high-energy data combined with the two sum rules~\eqref{eq:GinSumRules} significantly reduce the fit errors.

\begin{figure}[t]
    \centering
    \includegraphics[scale=0.5]{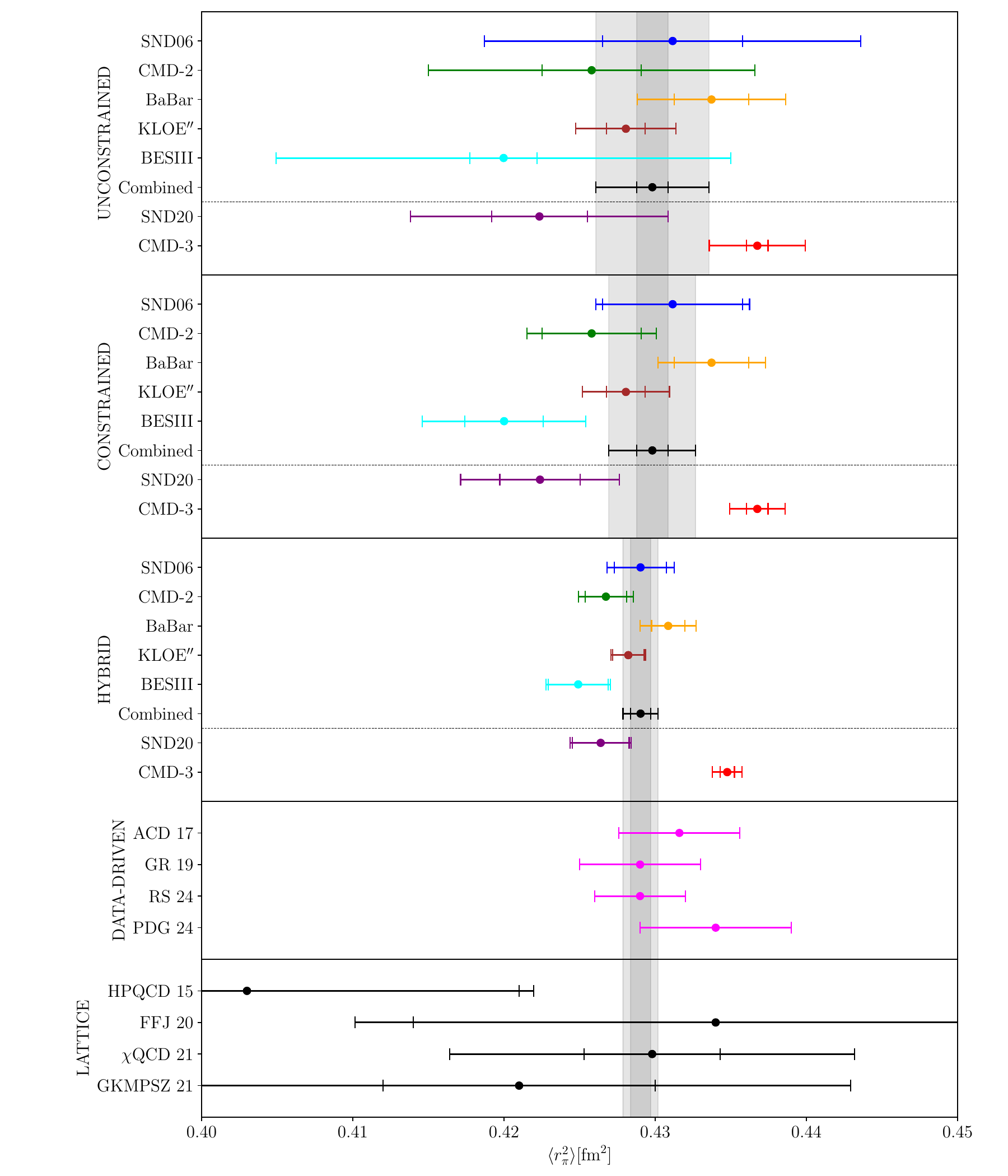}
    \caption{Values of $\< r_\pi^2\>$ (in $\text{fm}^2$) from the low-energy unconstrained and constrained fits and from the hybrid fits to single experiments. Error bars and combination are defined as in Fig.~\ref{fig:a_mu_1GeV_default_nozero}. Below, some other data-driven results from $e^+e^-$ data~\cite{Ananthanarayan:2017efc,RuizArriola:2024gwb} and $\tau$ data~\cite{Gonzalez-Solis:2019iod} as well as the current PDG value~\cite{ParticleDataGroup:2024cfk}, and some lattice results~\cite{Koponen:2015tkr,Feng:2019geu,Wang:2020nbf,Gao:2021xsm} are plotted for comparison.}
    \label{fig:r_pi2}
\end{figure}

The results are presented in Fig.~\ref{fig:r_pi2}, using symmetric errors. Interestingly, in the hybrid fits the BESIII, SND20, and CMD-3 values are shifted towards the middle. We interpret this as an effect of the combination with the BaBar data above $1.4\GeV$, which pulls the different results towards the BaBar value. The goodness of fit to CMD-3 is much reduced in the hybrid representation, indicating a tension of CMD-3 even with BaBar data at higher energies. For these reasons, as central results for the pion charge radius from individual experiments we prefer the low-energy constrained fits, which avoid any mixing of data sets. For the combined fit, we take the hybrid representation as central value, since the discussed issues do not apply here. These results are presented in Table~\ref{tab:r_pi2}. For the combined fit, we include the difference to the low-energy constrained value as an additional systematic effect. Our final result from a combined fit to all $e^+e^-$ experiments (apart from SND20 and CMD-3) as well as the NA7 data set reads
\begin{equation}
	\< r_\pi^2\>|^\text{comb} = 0.4290(7)_{\text{stat}}(9)_{\text{syst}}(9)_\text{BK}(8)_{\text{low-E}}\fm^2 = 0.4290(17)\fm^2 \, ,
\end{equation}
whereas the constrained fit to CMD-3 alone gives
\begin{equation}
	\< r_\pi^2 \>|^\text{CMD-3} = 0.4367(7)_\text{stat}(23)_\text{syst} \fm^2 = 0.4367(24) \fm^2 \, .
\end{equation}
A comparison with existing data-driven and lattice-QCD results is shown in Table~\ref{tab:r_pi2_comparison}. Our determination of $\<r_\pi^2\>$ is in good agreement with all the other results but more precise. The determination of $\<r_\pi^2\>$ is clearly limited by the tensions between the different $e^+e^-$ experiments. Conversely, the reduction of the systematic uncertainty renders the pion charge radius a relevant observable to discriminate between different $e^+e^-$ experiments. The significance of the tensions in $\<r_\pi^2\>$ between the fits to CMD-3 and the other experiments are listed in Table~\ref{tab:discrepancies_rpi2}. For the low-energy fits, they strongly depend on the choice whether to use symmetric or asymmetric errors, since the dominant uncertainty comes from the very asymmetric effect of the variation of $N$. This effect is very prominent in the unconstrained fits, but much reduced if we impose the absence of zeros. Although current uncertainties in lattice-QCD evaluations of $\<r_\pi^2\>$ are not yet at a competitive level (see Fig.~\ref{fig:r_pi2}), the pion charge radius offers an opportunity for future lattice-QCD computations to probe the present discrepancies, independently of full HVP evaluations.

\begin{table}[t]
    \centering
    \input{tables/r_pi2_vs_N}

    \caption{Comparison of the values of $\< r_\pi^2\>$ (in units of $\text{fm}^2$) in the unconstrained (without brackets) and constrained fits (with brackets) to single experiments for different values of $N$. Combination and errors are defined as in Table~\ref{tab:detailed_single_exp}. For large values of $N$, the charge radius starts to become sensitive to fit instabilities, reflected in large .}
    \label{tab:r_pi2_vs_N}
\end{table}

\begin{table}[t]
    \centering
    \input{tables/r_pi2}

    \caption{Values of $\< r_\pi^2\>$ in the unconstrained and constrained low-energy fits and hybrid fits. Combination and errors are defined as in Table~\ref{tab:detailed_single_exp}.}
    \label{tab:r_pi2}
\end{table}

\begin{table}[t]
    \centering
    \input{tables/r_pi2_comparison}
    \caption{Results for $\< r_\pi^2\>$ from our combined fits (all $e^+e^-$ experiments apart from SND20 and CMD-3, and the NA7 data set) and our fits to CMD-3, in comparison with other data-driven determinations ($e^+e^-$~\cite{Ananthanarayan:2017efc,Colangelo:2018mtw,RuizArriola:2024gwb}, $\tau$~\cite{Gonzalez-Solis:2019iod}, PDG~\cite{ParticleDataGroup:2024cfk}) and lattice-QCD results~\cite{Koponen:2015tkr,Feng:2019geu,Wang:2020nbf,Gao:2021xsm}.}
    \label{tab:r_pi2_comparison}
\end{table}

\begin{table}[t]
    \centering
    \input{tables/discrepancies_rpi2}
    \caption{Significance of the discrepancies in the pion charge radius $\< r_\pi^2\>$ between fits to CMD-3 and the other experiments, taking into account the correlations due to the systematics in the dispersive representation, as well as the $\chi^2$ inflation of the fit errors. The significances in curly brackets ignore the strong asymmetries in the systematic error due to the variation of $N$ (assigning the maximum variation symmetrically). The numbers without brackets treat the variation of $N$ asymmetrically. The significances in square brackets exclude the systematic effect due to the BaBar--KLOE tension.}
    \label{tab:discrepancies_rpi2}
\end{table}

\subsection{Effect of correlations}

In our fits, we take into account the full covariance matrices for the experimental data whenever this information is provided. see Ref.~\cite{Colangelo:2018mtw} for details.\footnote{The covariance matrix for the BESIII data~\cite{BESIII:2015equ} was corrected after problems were pointed out in Ref.~\cite{Colangelo:2018mtw}.}  In the case of CMD-3, the statistical uncertainties are taken as uncorrelated, whereas the systematic uncertainty, which to a large extent concerns the overall normalization, is taken as fully correlated between all data points. In order to avoid the D'Agostini bias~\cite{DAgostini:1993arp}, we implement the iterative method introduced by the NNPDF collaboration~\cite{Ball:2009qv}.

\begin{figure}[t]
    \centering
    \scriptsize
    \includegraphics[width=14cm]{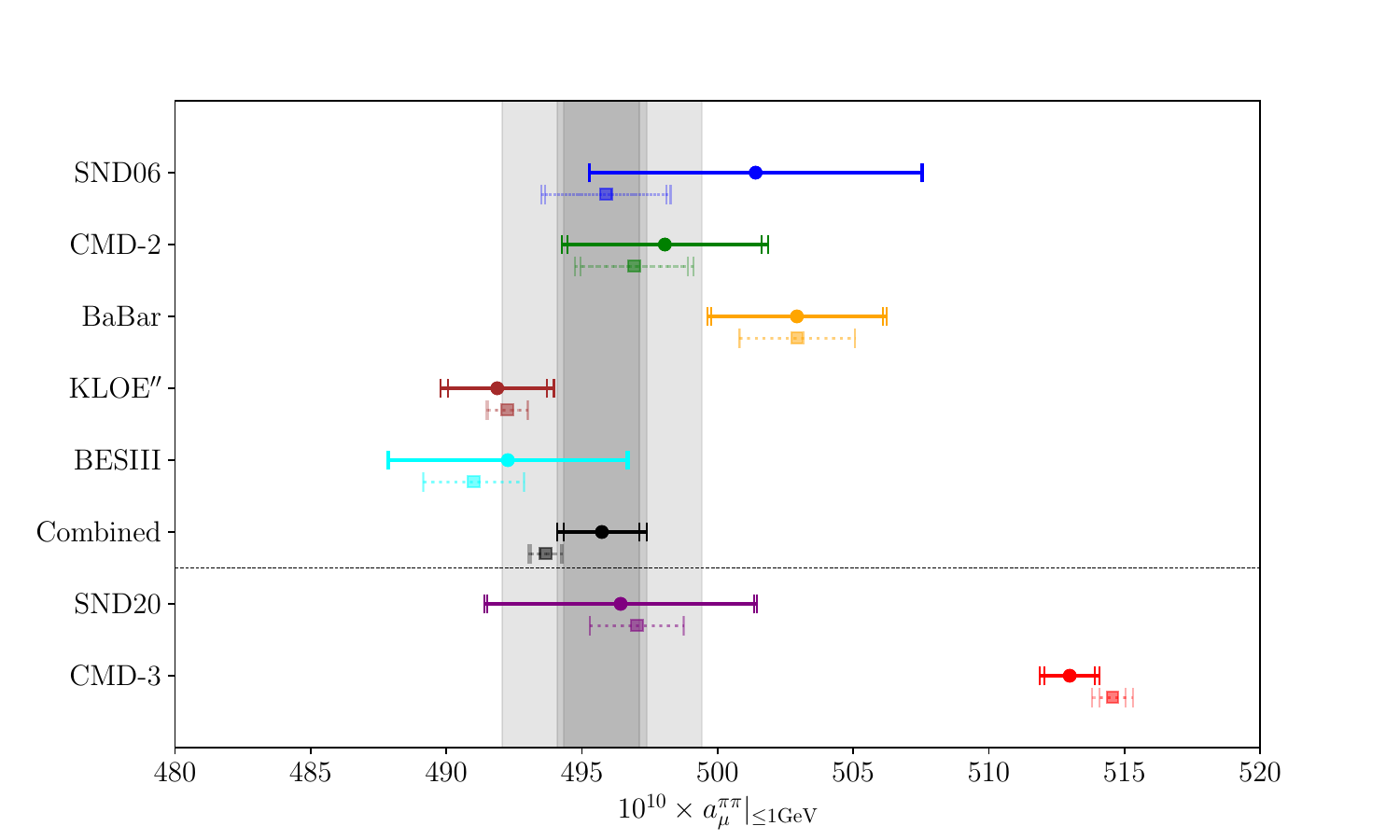}
    \caption{Result for $a_\mu^{\pi\pi}|_{\le1\GeV}$ from the constrained low-energy fits: the filled dots with solid error bars show our central result, while the transparent squares with dotted error bars are the results obtained by taking into account only diagonal systematic errors.}
    \label{fig:NoCorrelation}
\end{figure}

\begin{figure}[t]
    \centering
    \scriptsize
    \begin{tabular}{ccccc}
    \includegraphics[width=1.8cm]{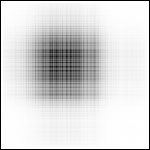} &
    \includegraphics[width=1.8cm]{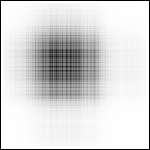} &
    \includegraphics[width=1.8cm]{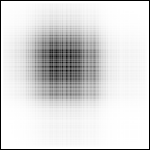} &
    \includegraphics[width=1.8cm]{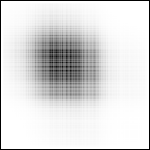} &
    \includegraphics[width=1.8cm]{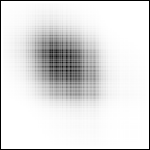}
    \\
    $\alpha=0$ &
    $\alpha=0.11$ &
    $\alpha=0.22$ &
    $\alpha=0.33$ &
    $\alpha=0.44$
    \\ \\
    \includegraphics[width=1.8cm]{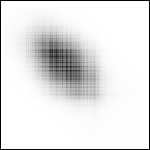} &
    \includegraphics[width=1.8cm]{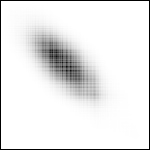} &
    \includegraphics[width=1.8cm]{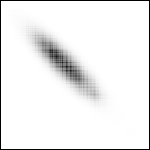} &
    \includegraphics[width=1.8cm]{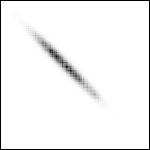} &
    \includegraphics[width=1.8cm]{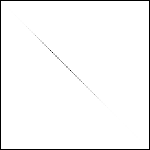}
    \\
    $\alpha=0.56$ &
    $\alpha=0.67$ &
    $\alpha=0.78$ &
    $\alpha=0.89$ &
    $\alpha=1.0$
    \end{tabular}
    \caption{Modified systematic covariance matrices for CMD-3 used to check the effect of reducing correlations.}
    \label{fig:Decorrelation}
\end{figure}

\begin{figure}[t]
    \centering
    \scriptsize
    \includegraphics[width=10cm]{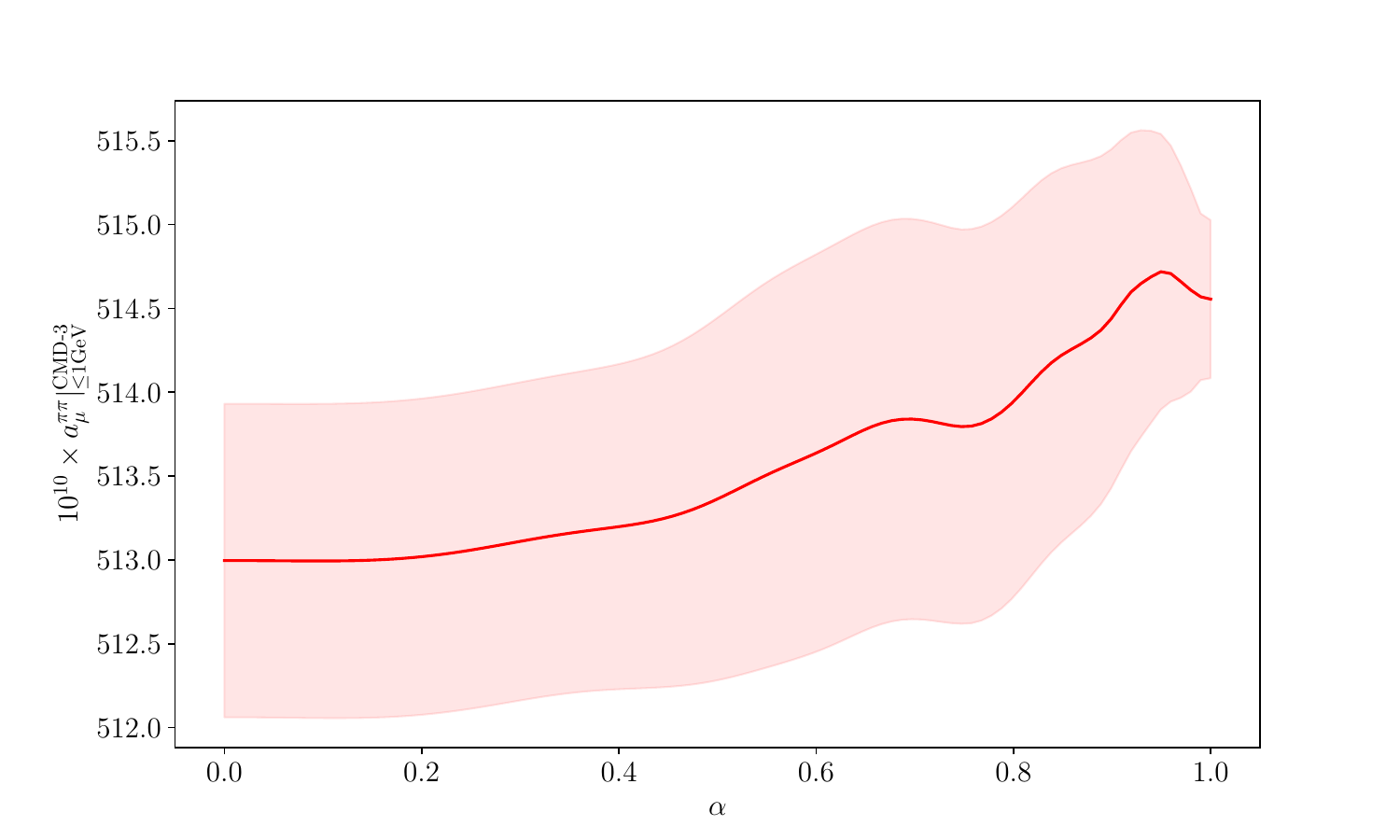}
    \caption{Result for $a_\mu^{\pi\pi}|_{\le1\GeV}$ in the constrained low-energy fit to CMD-3 including fit uncertainty, inflated by $\sqrt{\chi^2/\mathrm{dof}}$, as a function of the decorrelation parameter $\alpha$.}
    \label{fig:DecorrelationResult}
\end{figure}

The experimental determinations of covariances are subject to uncertainties themselves, see the discussion in Ref.~\cite{Aoyama:2020ynm}. However, since this information currently is not provided, we neglect it as an effect of second order in uncertainties. Although we take the position that all information on uncertainties and correlations should come from the experiments, we can still try to estimate the impact of the assumptions that went into our analysis. In particular, since CMD-3 does not provide full covariance matrices, in the following we will study the effect of relaxing the systematic correlations between the data points. Using only diagonal errors instead of fully correlated ones typically leads to significantly smaller uncertainties, but also affects the central values, as shown in Fig.~\ref{fig:NoCorrelation}. The dependence of central values and uncertainties on the correlations of the fit data however is not monotonous.\footnote{We thank B.~Malaescu for pointing this out and for suggesting to investigate the issue.} As a simple toy scenario, we assign systematic correlations to the 209 data points of the CMD-3 data set as a function of the distance in bin index number $i$ and $j$,
\begin{equation}
	\mathrm{Corr}_{ij} = \exp\left\{ - 10 \frac{ (i-j)^2 }{r(\alpha)^2} \right\} \, , \quad r(\alpha) = 209 \frac{1-\alpha}{\alpha} \, ,
\end{equation}
with parameter $\alpha$ scanned between $0$ and $1$. For $\alpha=0$, we recover full correlation of the systematic errors, as used for our central results, whereas for $\alpha=1$ only diagonal errors are retained. The resulting systematic covariance matrices are illustrated in Fig.~\ref{fig:Decorrelation}. In Fig.~\ref{fig:DecorrelationResult}, we show the result for $a_\mu^{\pi\pi}|_{\le1\GeV}$ in the constrained low-energy fit to CMD-3 as a function of the decorrelation parameter $\alpha$. As long as no explicit systematic correlation matrix is provided by the experiment, we take the case of fully correlated systematic errors as the most realistic scenario. The results for the toy example considered here suggest however that reduced correlations will invariably lead to even larger discrepancies. Therefore, we consider our current treatment as a conservative approach.

%% file: tables/detailed_single_exp.tex

\resizebox{\textwidth}{!}{
\begin{tabular}{lccllllll}
\toprule{} \\[-10pt]
		& $\chi^2$/dof		& $p$-value		& $\delta_1^1(s_0)$ [°]		& $\delta_1^1(s_1)$ [°]		& $M_{\omega}$ [MeV]		& $10^3\times\text{Re}\,\epsilon_{\omega}$		& $\delta_\epsilon$ [°] $\qquad\quad$		& $10^{10}\times a_\mu^{\pi\pi}|_{\leq1\text{GeV}}$ \\[3pt]
\midrule{}
SND06		& $1.09$	& $33.3\%$	& $110.8(4)(6)$		& $166.3(0.3)(2.0)$		& $782.11(33)(1)$		& $2.03(5)(1)$		& $8.6(2.3)(0.3)$		& $497.8(6.1)(4.1)$ \\{}
[SND06]		& $1.17$	& $21.8\%$	& $110.3(4)(4)$		& $166.1(0.3)(1.9)$		& $782.12(34)(2)$		& $2.04(5)(1)$		& $8.8(2.4)(0.4)$		& $501.5(6.1)(0.9)$ \\{}
(SND06)		& $0.99$	& $50.0\%$	& $110.5(3)(3)$		& $166.3(0.3)(1.8)$		& $782.15(32)(1)$		& $2.03(4)(1)$		& $9.3(2.2)(0.3)$		& $499.1(4.7)(2.1)$ \\[5pt]{}
%
CMD-2		& $1.01$	& $45.8\%$	& $111.0(5)(9)$		& $167.2(0.4)(2.0)$		& $782.66(33)(3)$		& $1.90(6)(3)$		& $11.5(3.1)(0.7)$		& $495.8(3.7)(3.4)$ \\{}
[CMD-2]		& $1.05$	& $36.1\%$	& $110.3(5)(4)$		& $166.9(0.4)(2.0)$		& $782.65(34)(4)$		& $1.91(6)(1)$		& $11.4(3.2)(0.8)$		& $498.2(3.6)(1.3)$ \\{}
(CMD-2)		& $0.96$	& $59.2\%$	& $110.4(4)(2)$		& $166.9(0.3)(1.8)$		& $782.65(32)(10)$		& $1.91(6)(1)$		& $11.4(3.0)(0.4)$		& $498.4(3.2)(1.1)$ \\[5pt]{}
%
BaBar		& $1.17$	& $3.0\%$		& $110.2(3)(5)$		& $165.5(0.2)(2.0)$		& $781.89(18)(2)$		& $2.06(4)(1)$		& $0.3(1.9)(0.6)$		& $501.8(3.3)(1.6)$ \\{}
[BaBar]		& $1.17$	& $2.9\%$		& $110.0(3)(4)$		& $165.5(0.2)(2.0)$		& $781.91(18)(2)$		& $2.07(4)(1)$		& $0.8(1.9)(0.4)$		& $502.9(3.2)(0.6)$ \\{}
(BaBar)		& $1.12$	& $6.7\%$		& $110.2(2)(3)$		& $165.7(0.2)(1.8)$		& $781.91(18)(1)$		& $2.04(3)(2)$		& $0.5(1.9)(0.4)$		& $500.2(2.5)(2.6)$ \\[5pt]{}
%
KLOE$''$		& $1.13$	& $10.6\%$	& $110.4(2)(5)$		& $165.9(0.2)(2.0)$		& $782.45(24)(5)$		& $1.96(4)(2)$		& $6.0(1.7)(0.4)$		& $490.9(2.0)(1.7)$ \\{}
[KLOE$''$]	& $1.13$	& $10.5\%$	& $110.3(1)(4)$		& $165.8(0.1)(1.9)$		& $782.40(23)(5)$		& $1.98(4)(1)$		& $5.8(1.7)(0.4)$		& $491.8(1.8)(0.9)$ \\{}
(KLOE$''$)	& $1.09$	& $17.8\%$	& $110.3(1)(3)$		& $165.8(0.1)(1.9)$		& $782.41(23)(4)$		& $1.98(4)(1)$		& $5.8(1.7)(0.5)$		& $492.0(1.7)(0.7)$ \\[5pt]{}
%
BESIII		& $1.01$	& $44.5\%$	& $111.3(0.9)(1.2)$	& $167.4(0.5)(2.0)$		& $783.07(61)(2)$		& $2.03(19)(6)$		& $17.8(6.9)(1.2)$		& $490.4(4.5)(2.7)$ \\{}
[BESIII]		& $1.02$	& $42.7\%$	& $110.7(5)(6)$		& $167.3(0.5)(2.0)$		& $783.09(60)(1)$		& $2.07(19)(3)$		& $18.7(6.7)(0.3)$		& $492.3(4.4)(0.9)$ \\{}
(BESIII)		& $0.94$	& $62.0\%$	& $110.3(5)(5)$		& $167.1(0.4)(1.9)$		& $783.04(58)(4)$		& $2.11(18)(2)$		& $17.8(6.3)(1.1)$		& $494.1(4.0)(2.4)$ \\[5pt]{}
%
SND20		& $1.88$	& $0.4\%$		& $110.9(6)(8)$		& $167.0(0.3)(2.0)$		& $782.34(28)(6)$		& $2.07(5)(1)$		& $9.9(2.4)(1.3)$		& $495.1(5.3)(2.5)$ \\{}
[SND20]		& $1.85$	& $0.4\%$		& $110.5(3)(5)$		& $167.0(0.3)(2.0)$		& $782.38(27)(3)$		& $2.07(5)(1)$		& $10.6(2.2)(0.7)$		& $496.5(4.9)(1.3)$ \\{}
(SND20)		& $1.39$	& $3.9\%$		& $110.4(4)(3)$		& $166.9(0.3)(1.9)$		& $782.40(24)(2)$		& $2.08(4)(1)$		& $10.8(2.0)(0.4)$		& $497.4(4.1)(1.2)$ \\[5pt]{}
%
CMD-3		& $1.09$	& $19.7\%$	& $110.7(1)(4)$		& $166.2(0.1)(2.0)$		& $782.33(6)(1)$		& $2.08(1)(1)$		& $7.4(4)(3)$		& $513.7(1.1)(0.7)$ \\{}
[CMD-3]		& $1.10$	& $18.3\%$	& $110.8(1)(4)$		& $166.2(0.1)(2.0)$		& $782.32(6)(2)$		& $2.08(1)(1)$		& $7.1(4)(3)$		& $513.0(0.9)(0.8)$ \\{}
(CMD-3)		& $1.24$	& $0.9\%$		& $111.0(1)(3)$		& $166.3(0.1)(1.8)$		& $782.29(6)(4)$		& $2.08(1)(1)$		& $7.1(4)(2)$		& $511.2(1.4)(1.6)$ \\[2pt]
\midrule{}
Combination	& $1.21$	& $1.4\times 10^{-4}$		& $110.3(1)(5)$		& $165.8(0.1)(2.0)$		& $782.07(12)(1)$		& $1.99(2)(1)$		& $3.8(9)(4)$		& \hspace{-0.5cm} $494.8(1.5)(1.4)(3.4)$ \\{}
[Combination]	& $1.21$	& $1.1\times 10^{-4}$		& $110.2(1)(4)$		& $165.8(0.1)(2.0)$		& $782.07(12)(1)$		& $2.00(2)(1)$		& $4.0(9)(3)$		& \hspace{-0.5cm} $495.7(1.4)(0.8)(3.3)$ \\{}
(Combination)	& $1.19$	& $2.4\times 10^{-4}$		& $110.2(1)(3)$		& $165.8(0.1)(1.8)$		& $782.09(12)(2)$		& $1.99(2)(1)$		& $4.2(9)(2)$		& \hspace{-0.5cm} $494.9(1.2)(1.7)(2.7)$ \\[2pt]
\bottomrule
\end{tabular}
}

%% file: tables/a_mu_vs_N.tex

\resizebox{\textwidth}{!}{
\begin{tabular}{lllllllll}
\toprule{} \\[-10pt]
$10^{10}\times a_\mu^{\pi\pi}|_{\leq1\text{GeV}}$ & & & & & & & & \\ [5pt]
$N-1$ & 1 & 2 & 3 & 4 & 5 & 6 & 7 & Central \\[3pt]
\midrule{} \\[-10pt]
SND06 & 			$500.8(3.6)$	&	$501.9(4.8)$	&	$501.4(6.1)$	&	$497.9(6.1)$	&	$497.5(6.2)$	&	$496.0(6.2)$	&	$493.2(6.4)$	&	$497.8(6.1)(4.1)$ \\
$[\text{SND06}]$ & 	$500.8(3.6)$	&	$501.9(4.8)$	&	$501.4(6.1)$	&	$501.2(6.1)$	&	$501.4(6.0)$	&	$499.8(6.7)$	&	$499.5(6.9)$	&	$501.5(6.1)(0.9)$ \\[5pt]
CMD-2 & 			$499.7(3.0)$	&	$499.2(3.2)$	&	$498.0(3.6)$	&	$495.8(3.7)$	&	$495.9(3.7)$	&	$495.2(4.0)$	&	$495.3(4.0)$ 	&	$495.8(3.7)(3.4)$ \\
$[\text{CMD-2}]$ & 	$499.7(3.0)$	&	$499.2(3.2)$	&	$498.0(3.6)$	&	$497.9(3.6)$	&	$498.1(3.9)$	&	$496.5(3.8)$	&	$496.5(4.1)$	&	$498.2(3.6)(1.3)$ \\[5pt]
BaBar & 			$499.8(2.4)$	&	$503.5(3.2)$	&	$503.0(3.2)$	&	$501.9(3.3)$	&	$501.6(3.5)$	&	$500.9(3.6)$	&	$500.8(3.6)$	&	$501.8(3.3)(1.6)$ \\
$[\text{BaBar}]$ & 	$499.8(2.4)$	&	$503.5(3.2)$	&	$503.0(3.2)$	&	$502.7(3.2)$	&	$502.4(3.2)$	&	$502.3(3.3)$	&	$502.0(3.4)$	&	$502.9(3.2)(0.6)$ \\[5pt]
KLOE$''$ & 		$491.1(1.4)$	&	$492.5(1.8)$	&	$491.8(1.8)$	&	$490.9(2.0)$	&	$491.3(2.2)$	&	$491.2(2.2)$	&	$491.6(2.3)$	&	$490.9(2.0)(1.7)$ \\
$[\text{KLOE$''$}]$ & $491.1(1.4)$	&	$492.5(1.8)$	&	$491.8(1.8)$	&	$491.4(1.8)$	&	$491.3(2.1)$	&	$491.2(2.1)$	&	$491.2(2.2)$	&	$491.8(1.8)(0.9)$ \\[5pt]
BESIII & 			$493.3(4.3)$	&	$493.0(4.4)$	&	$492.3(4.4)$	&	$490.4(4.5)$	&	$490.0(4.6)$	&	$488.9(4.6)$	&	$485.1(4.6)$	&	$490.4(4.5)(2.7)$ \\
$[\text{BESIII}]$ & 	$493.3(4.3)$	&	$493.0(4.4)$	&	$492.3(4.4)$	&	$492.3(4.4)$	&	$492.3(6.1)$	&	$491.5(4.5)$	&	$491.5(4.8)$	&	$492.3(4.4)(0.9)$ \\[5pt]
SND20 & 			$497.9(4.8)$	&	$497.5(4.9)$	&	$496.4(4.9)$	&	$495.1(5.3)$	&	$495.0(5.4)$	&	$497.1(5.5)$	&	$500.7(7.6)$	&	$495.1(5.3)(2.5)$ \\
$[\text{SND20}]$ & 	$497.9(4.8)$	&	$497.5(4.9)$	&	$496.4(4.9)$	&	$496.4(5.0)$	&	$496.4(5.1)$	&	$495.5(6.0)$	&	$495.5(8.7)$	&	$496.5(4.9)(1.3)$ \\[5pt]
CMD-3 & 			$509.8(0.9)$	&	$513.3(0.9)$	&	$513.0(0.9)$	&	$513.7(1.1)$	&	$513.7(1.1)$	&	$513.8(1.1)$	&	$513.8(1.1)$	&	$513.7(1.1)(0.7)$ \\
$[\text{CMD-3}]$ & 	$509.8(0.9)$	&	$513.3(0.9)$	&	$513.0(0.9)$	&	$513.7(1.1)$	&	$513.6(1.1)$	&	$513.8(1.1)$	&	$513.8(1.1)$	&	$513.0(0.9)(0.8)$ \\[5pt]
\midrule{} \\[-10pt]
Combination & 			$494.1(1.1)$	&	$496.2(1.4)$	&	$495.7(1.4)$	&	$494.8(1.5)$	&	$494.9(1.5)$	&	$494.6(1.5)$	&	$494.6(1.5)$	&	$494.8(1.5)(1.4)(3.4)$ \\
$[\text{Combination}]$ & 	$494.1(1.1)$	&	$496.2(1.4)$	&	$495.7(1.4)$	&	$495.4(1.4)$	&	$495.0(1.3)$	&	$495.0(1.5)$	&	$495.0(1.5)$	&	$495.7(1.4)(0.8)(3.3)$ \\[5pt]
\bottomrule
\end{tabular}
}

%% file: tables/discrepancies_amu.tex

\resizebox{\textwidth}{!}{
\begin{tabular}{l|ccc|cccc|c}
\toprule
\multirow{2}{*}{Discrepancy with CMD-3} & & $a_\mu^{\pi\pi}|_{\leq1\text{GeV}}$ & & SD window & int window & LD window & VLD window & $a_\mu^{\pi\pi}$ \\[5pt]
 & Unconstrained & Constrained & Hybrid & \multicolumn{4}{c|}{Constrained} & Hybrid  \\[2pt]
\midrule{} \\[-10pt]
SND06 &		$2.0\sigma$ & $1.8\sigma$ & $2.5\sigma$ & $1.8\sigma$ & $1.8\sigma$ & $1.8\sigma$ & $1.6\sigma$ & $2.6\sigma$ \\
CMD-2 &		$3.3\sigma$ & $3.7\sigma$ & $3.9\sigma$ & $3.4\sigma$ & $3.5\sigma$ & $3.8\sigma$ & $3.4\sigma$ & $4.1\sigma$ \\
BaBar &		$2.9\sigma$ & $2.8\sigma$ & $3.8\sigma$ & $2.9\sigma$ & $3.0\sigma$ & $2.8\sigma$ & $1.8\sigma$ & $3.7\sigma$ \\
KLOE$''$ &	$7.4\sigma$ & $8.9\sigma$ & $8.3\sigma$ & $9.9\sigma$ & $9.7\sigma$ & $8.5\sigma$ & $5.3\sigma$ & $7.1\sigma$ \\
BESIII &		$4.2\sigma$ & $4.5\sigma$ & $3.4\sigma$ & $3.8\sigma$ & $4.1\sigma$ & $4.6\sigma$ & $4.2\sigma$ & $3.4\sigma$ \\
SND20 &		$3.0\sigma$ & $3.2\sigma$ & $3.2\sigma$ & $2.6\sigma$ & $2.8\sigma$ & $3.3\sigma$ & $3.2\sigma$ & $3.2\sigma$ \\[2pt]
\midrule{} \\[-10pt]
Combination & $4.4\sigma$ [$7.3\sigma$] & $4.4\sigma$ [$8.1\sigma$] & $5.4\sigma$ [$9.4\sigma$] & $4.1\sigma$ [$9.3\sigma$] & $4.3\sigma$ [$9.1\sigma$] & $4.4\sigma$ [$7.6\sigma$] & $3.3\sigma$ [$4.2\sigma$] & $5.3\sigma$ [$9.3\sigma$] \\[2pt]
\bottomrule
\end{tabular}
}

%% file: tables/a_mu_ranges.tex

\resizebox{\textwidth}{!}{
\begin{tabular}{llllllll}
\toprule{} \\[-10pt]
$10^{10}\times{}$ & $a_\mu^{\pi\pi}|_{\leq0.63\GeV}$ & $a_\mu^{\pi\pi}|_{[0.6;0.9]\GeV}$ & $a_\mu^{\pi\pi}|_{\leq1\GeV}$ & $a_\mu^{\pi\pi}|_{\leq1.8\GeV}$ & $a_\mu^{\pi\pi}|_{[0.305;1.937]\GeV}$ & $a_\mu^{\pi\pi}|_{\leq1.937\GeV}$ & $a_\mu^{\pi\pi}|_{\leq3\GeV}$ \\[2pt]
\midrule{} \\[-10pt]
SND06 &		$133.7(1.3)(0.3)$ & $375.0(5.1)(0.7)$ & $501.5(6.2)(0.9)$ 		&	$508.5(4.9)(2.0)$ & $507.8(4.9)(2.0)$ & $508.7(4.9)(2.0)$ & $508.8(4.9)(2.0)$ \\
CMD-2 &		$131.9(1.0)(0.6)$ & $373.2(2.9)(0.8)$ & $498.2(3.6)(1.3)$ 		&	$507.5(3.3)(1.0)$ & $506.8(3.2)(1.0)$ & $507.7(3.3)(1.0)$ & $507.7(3.3)(1.0)$ \\
BaBar &		$134.5(0.7)(0.3)$ & $375.7(2.7)(0.4)$ & $502.9(3.2)(0.6)$ 		&	$510.2(2.5)(2.6)$ & $509.4(2.5)(2.6)$ & $510.3(2.5)(2.6)$ & $510.4(2.5)(2.6)$ \\
KLOE$''$ &	$132.8(0.4)(0.4)$ & $366.6(1.5)(0.6)$ & $491.8(1.8)(0.9)$ 		&	$501.7(1.9)(0.8)$ & $500.9(1.9)(0.8)$ & $501.8(1.9)(0.8)$ & $501.9(1.9)(0.8)$ \\
BESIII &		$130.5(0.9)(0.8)$ & $368.6(3.7)(0.6)$ & $492.3(4.4)(0.9)$ 		&	$503.0(4.4)(2.8)$ & $502.2(4.4)(2.8)$ & $503.1(4.4)(2.8)$ & $503.2(4.4)(2.8)$ \\
SND20 &		$131.3(0.9)(0.9)$ & $371.9(4.1)(1.0)$ & $496.5(5.0)(1.3)$ 		&	$506.4(4.4)(1.1)$ & $505.7(4.4)(1.1)$ & $506.6(4.4)(1.1)$ & $506.6(4.4)(1.1)$ \\
CMD-3 &		$135.6(0.2)(0.3)$ & $384.8(0.8)(0.5)$ & $513.0(0.9)(0.8)$ 		&	$521.6(0.9)(1.5)$ & $520.8(0.9)(1.5)$ & $521.7(0.9)(1.5)$ & $521.8(0.9)(1.5)$  \\[2pt]
\midrule{} \\[-10pt]
Combination & 	$133.3(0.3)(0.4)(0.5)$ & $369.9(1.2)(0.5)(2.7)$ & $495.7(1.4)(0.8)(3.3)$ & 	$504.5(1.2)(1.6)(2.9)$ & $503.8(1.2)(1.6)(2.9)$ & $504.7(1.2)(1.6)(2.9)$ & $504.7(1.2)(1.6)(2.9)$ \\
		 &	$133.3(0.7)$ & $369.9(3.0)$ & $495.7(3.7)$ & 							$504.5(3.5)$ 	& $503.8(3.5)$ & $504.7(3.5)$ & $504.7(3.5)$ \\[2pt]
\midrule{} \\[-10pt]
ACD 18~\cite{Ananthanarayan:2018nyx} & 132.9(0.8) & -- & -- & -- & -- & -- & --\\
Ref. \cite{Colangelo:2022vok} & -- & -- & $494.3(3.6)$ & -- & -- & -- & -- \\
DHMZ 19~\cite{Davier:2019can,Aoyama:2020ynm} & 132.9(0.5)(0.6) & $371.5(1.5)(2.3)$ & 497.4(1.8)(3.1) & $507.9(0.8)(3.2)(0.6)$ & -- & -- & -- \\
KNT 19~\cite{Keshavarzi:2019abf,Aoyama:2020ynm} & 131.2(1.0) & 369.8(1.3) & $493.8(1.9)$ & $504.2(1.9)$ & $503.5(1.9)$ & $504.3(1.9)$  & -- \\[2pt]
\bottomrule
\end{tabular}
}

%% file: tables/a_mu_windows.tex

\resizebox{0.75\textwidth}{!}{
\begin{tabular}{lllll}
\toprule{} \\[-10pt]
$10^{10}\times{}a_\mu^{\pi\pi}|_{\leq1\GeV}$ & SD window & int window & LD window & VLD window  \\[2pt]
\midrule{} \\[-10pt]
SND06		& $13.9(2)(1)$ & $140.5(1.8)(0.2)$ & $347.1(4.2)(0.6)$ & $26.7(2)(1)$ \\
CMD-2		& $13.9(1)(1)$ & $139.8(1.1)(0.3)$ & $344.5(2.5)(1.0)$ & $26.4(2)(1)$ \\
BaBar		& $14.0(1)(1)$ & $140.8(1.0)(0.2)$ & $348.2(2.2)(0.5)$ & $26.8(1)(1)$ \\
KLOE$''$		& $13.6(1)(1)$ & $137.3(0.5)(0.2)$ & $340.9(1.2)(0.6)$ & $26.4(1)(1)$ \\
BESIII		& $13.7(1)(1)$ & $138.1(1.4)(0.1)$ & $340.5(2.9)(0.8)$ & $26.1(2)(1)$ \\
SND20		& $13.9(2)(1)$ & $139.4(1.5)(0.2)$ & $343.2(3.3)(1.1)$ & $26.3(2)(1)$ \\
CMD-3		& $14.3(1)(1)$ & $143.8(0.3)(0.2)$ & $354.8(0.6)(0.6)$ & $27.1(1)(1)$ \\[2pt]
\midrule{} \\[-10pt]
Combination	& $13.7(1)(1)(1)$ & 	$138.5(0.4)(0.2)(1.1)$ & 	$343.4(0.9)(0.6)(2.1)$ & 	$26.5(1)(1)(1)$ \\
			& $13.7(1)$ & 		$138.5(1.2)$ & 			$343.4(2.6)$ & 			$26.5(1)$ \\[2pt]
\midrule{} \\[-10pt]
Ref.~\cite{Colangelo:2022vok} &	$13.7(1)$ & $138.3(1.2)$ & $342.3(2.3)$ & -- \\
\midrule
\midrule{} \\[-10pt]
$10^{10}\times{}a_\mu^{\pi\pi}$ \\[2pt]
\midrule{} \\[-10pt]
SND06		& $15.0(2)(1)$	&	$145.2(1.6)(0.5)$	&	$348.5(3.2)(1.4)$	&	$26.5(2)(1)$	\\
CMD-2		& $15.0(1)(1)$	&	$145.1(1.0)(0.2)$	&	$347.6(2.2)(0.8)$	&	$26.4(1)(1)$	\\
BaBar		& $15.1(1)(1)$	&	$145.7(0.8)(0.7)$	&	$349.5(1.7)(1.8)$	&	$26.6(1)(1)$	\\
KLOE$''$		& $14.8(1)(1)$	&	$142.9(0.6)(0.4)$	&	$344.2(1.2)(0.5)$	&	$26.4(1)(1)$	\\
BESIII		& $14.8(2)(2)$	&	$143.6(1.5)(1.0)$	&	$344.8(2.8)(1.6)$	&	$26.3(1)(1)$	\\
SND20		& $14.9(2)(1)$	&	$144.7(1.5)(0.3)$	&	$347.0(2.8)(0.8)$	&	$26.4(1)(1)$	\\
CMD-3		& $15.5(1)(1)$	&	$149.3(0.3)(0.4)$	&	$357.1(0.6)(1.1)$	&	$27.0(1)(1)$	\\[2pt]
\midrule{} \\[-10pt]
Combination	& $14.9(1)(1)(1)$	&	$143.8(0.4)(0.4)(1.0)$	&	$346.0(0.8)(1.1)(1.7)$	&	$26.5(1)(1)(1)$	\\
			& $14.9(1)$ 		& 	$143.8(1.2)$ 			& 	$346.0(2.2)$			&	$26.5(1)$	  \\[2pt]
\bottomrule
\end{tabular}
}

%% file: tables/r_pi2_vs_N.tex

\resizebox{0.9\textwidth}{!}{
\begin{tabular}{llllllll}
\toprule{} \\[-10pt]
$\langle r_\pi^2\rangle$ $[\fm^2]$ & & & & & & & \\ [5pt]
$N-1$ & 1 & 2 & 3 & 4 & 5 & 6 & 7 \\[2pt]
\midrule{} \\[-10pt]
SND06				&	$0.4311(12)$	&	$0.4316(20)$	&	$0.4311(46)$	&	$0.4216(58)$	&	$0.4203(61)$	&	$0.4125(72)$	&	$0.3942(154)$ \\
$[\text{SND06}]$		&	$0.4311(12)$	&	$0.4316(20)$	&	$0.4311(46)$	&	$0.4299(43)$	&	$0.4293(40)$	&	$0.4261(138)$	&	$0.4236(100)$ \\[5pt]
CMD-2				&	$0.4288(12)$	&	$0.4282(15)$	&	$0.4258(38)$	&	$0.4158(52)$	&	$0.4162(55)$	&	$0.4095(125)$	&	$0.4110(147)$ \\
$[\text{CMD-2}]$		&	$0.4288(12)$	&	$0.4282(15)$	&	$0.4258(38)$	&	$0.4248(34)$	&	$0.4246(126)$	&	$0.4196(53)$	&	$0.4193(231)$ \\[5pt]
BaBar				&	$0.4316(9)$	&	$0.4348(21)$	&	$0.4338(24)$	&	$0.4301(30)$	&	$0.4289(55)$	&	$0.4226(107)$	&	$0.4224(98)$ \\
$[\text{BaBar}]$			&	$0.4316(9)$	&	$0.4348(21)$	&	$0.4338(24)$	&	$0.4328(22)$	&	$0.4318(21)$	&	$0.4315(34)$	&	$0.4305(41)$ \\[5pt]
KLOE$''$				&	$0.4282(4)$	&	$0.4295(12)$	&	$0.4280(14)$	&	$0.4253(30)$	&	$0.4275(50)$	&	$0.4271(62)$	&	$0.4305(79)$ \\
$[\text{KLOE$''$}]$		&	$0.4282(4)$	&	$0.4295(12)$	&	$0.4280(14)$	&	$0.4266(13)$	&	$0.4270(36)$	&	$0.4270(51)$	&	$0.4270(53)$ \\[5pt]
BESIII				&	$0.4254(22)$	&	$0.4246(25)$	&	$0.4200(26)$	&	$0.4090(89)$	&	$0.4059(100)$	&	$0.3952(109)$	&	$0.3593(121)$ \\
$[\text{BESIII}]$		&	$0.4254(22)$	&	$0.4246(25)$	&	$0.4200(26)$	&	$0.4200(26)$	&	$0.4200(319)$	&	$0.4163(27)$	&	$0.4162(57)$ \\[5pt]
SND20				&	$0.4275(23)$	&	$0.4269(26)$	&	$0.4224(27)$	&	$0.4161(93)$	&	$0.4168(97)$	&	$0.4318(95)$	&	$0.4603(375)$ \\
$[\text{SND20}]$		&	$0.4275(23)$	&	$0.4269(26)$	&	$0.4224(27)$	&	$0.4224(28)$	&	$0.4222(249)$	&	$0.4187(134)$	&	$0.4185(146)$ \\[5pt]
CMD-3				&	$0.4338(3)$	&	$0.4375(6)$	&	$0.4368(7)$	&	$0.4386(15)$	&	$0.4395(17)$	&	$0.4399(33)$	&	$0.4407(40)$ \\
$[\text{CMD-3}]$		&	$0.4338(3)$	&	$0.4375(6)$	&	$0.4368(7)$	&	$0.4386(15)$	&	$0.4388(15)$	&	$0.4398(17)$	&	$0.4397(26)$ \\[2pt]
\midrule{} \\[-10pt]
Combination			&	$0.4291(4)$	&	$0.4310(9)$	&	$0.4298(10)$	&	$0.4267(21)$	&	$0.4275(25)$	&	$0.4234(32)$	&	$0.4233(44)$ \\
$[\text{Combination}]$	&	$0.4291(4)$	&	$0.4310(9)$	&	$0.4298(10)$	&	$0.4285(10)$	&	$0.4277(12)$	&	$0.4275(23)$	&	$0.4276(20)$ \\[2pt]
\bottomrule
\end{tabular}
}

%% file: tables/r_pi2.tex

\resizebox{0.6\textwidth}{!}{%
\begin{tabular}{llll}
\toprule{} \\[-10pt]
$\langle r_\pi^2\rangle$ [$\fm^2$]	&	Unconstrained	&	Constrained	&	Hybrid  \\[2pt]
\midrule{} \\[-10pt]
SND06	&	$0.4313(47)(109)$	&	$0.4313(47)(22)$	&	$0.4285(27)(12)$ \\
CMD-2	&	$0.4261(38)(101)$	&	$0.4261(38)(28)$	&	$0.4261(25)(10)$ \\
BaBar	&	$0.4337(24)(50)$	&	$0.4337(24)(23)$	&	$0.4308(11)(15)$ \\
KLOE$''$	&	$0.4280(14)(31)$	&	$0.4280(14)(20)$	&	$0.4281(10)(8)$ \\
BESIII	&	$0.4201(26)(149)$	&	$0.4201(26)(53)$	&	$0.4244(26)(19)$ \\
SND20	&	$0.4225(27)(80)$	&	$0.4225(27)(53)$	&	$0.4256(32)(9)$ \\
CMD-3	&	$0.4367(7)(30)$	&	$0.4367(7)(23)$	&	$0.4347(5)(8)$  \\[2pt]
\midrule{} \\[-10pt]
Combination	&	$0.4298(10)(34)(18)$	&	$0.4298(10)(25)(18)$	&	$0.4290(7)(9)(9)$ \\
				&	$0.4298(40)$	&	$0.4298(32)$	&	$0.4290(15)$ \\[2pt]
\bottomrule
\end{tabular}
}

%% file: tables/r_pi2_comparison.tex

\resizebox{0.55\textwidth}{!}{%
\begin{tabular}{lll}
\toprule{} \\[-10pt]
    $\< r_\pi^2\>$ [$\fm^2$] & & \\[2pt]
\midrule{} \\[-10pt]
    \multirow{3}{*}{This analysis, combination} & Unconstrained & $0.4298(40)$ \\
     & Constrained & $0.4298(32)$ \\
    & Hybrid & $0.4290(15)$ \\[2pt]
\midrule{} \\[-10pt]
    \multirow{3}{*}{This analysis, CMD-3} & Unconstrained & $0.4367(31)$ \\
     & Constrained & $0.4367(24)$ \\
    & Hybrid & $0.4347(9)$ \\[2pt]
\midrule{} \\[-10pt]
    \multirow{4}{*}{Data-driven}
     & ACD 17 \cite{Ananthanarayan:2017efc} & $0.432(4)$ \\
     & CHS 19~\cite{Colangelo:2018mtw} & $0.429(4)$ \\
     & GR 19~\cite{Gonzalez-Solis:2019iod} & $0.439(3)$ \\
     & RS 24~\cite{RuizArriola:2024gwb} & $0.429(4)$ \\
     & PDG 24~\cite{ParticleDataGroup:2024cfk} & $0.434(5)$ \\[2pt]
\midrule{} \\[-10pt]
    \multirow{4}{*}{Lattice} & HPQCD~15~\cite{Koponen:2015tkr} & $0.403(18)(6)$ \\
     & FFJ 20 \cite{Feng:2019geu} & $0.434(20)(13)$ \\
     & $\chi$QCD 21~\cite{Wang:2020nbf} & $0.430(5)(13)$ \\
     & GKMPSZ 21 \cite{Gao:2021xsm} & $0.421(9)(20)$ \\[2pt]
\bottomrule
\end{tabular}
}

%% file: tables/discrepancies_rpi2.tex

\resizebox{0.7\textwidth}{!}{%
\begin{tabular}{lccc}
\toprule{} \\[-10pt]
Discrepancy in $\< r_\pi^2 \>$ with CMD-3	&	Unconstrained	&	Constrained	&	Hybrid  \\[2pt]
\midrule{} \\[-10pt]
SND06	&	$\{0.4\sigma\}$ $1.1\sigma$	&	$\{0.9\sigma\}$ $1.1\sigma$	&	$2.2\sigma$ \\
CMD-2	&	$\{0.8\sigma\}$ $2.2\sigma$	&	$\{2.0\sigma\}$ $2.2\sigma$	&	$3.3\sigma$ \\
BaBar	&	$\{0.4\sigma\}$ $1.2\sigma$	&	$\{0.6\sigma\}$ $1.2\sigma$	&	$2.7\sigma$ \\
KLOE$''$	&	$\{1.8\sigma\}$ $4.8\sigma$	&	$\{2.4\sigma\}$ $4.9\sigma$	&	$4.7\sigma$ \\
BESIII	&	$\{0.9\sigma\}$ $2.2\sigma$	&	$\{2.5\sigma\}$ $2.6\sigma$	&	$3.1\sigma$ \\
SND20	&	$\{1.3\sigma\}$ $1.9\sigma$	&	$\{2.1\sigma\}$ $2.1\sigma$	&	$2.8\sigma$ \\[2pt]
\midrule{} \\[-10pt]
Combination	&	$\{1.3\sigma\}$ $3.1\sigma$ $[5.0\sigma]$	&	$\{1.5\sigma\}$ $3.1\sigma$ $[5.0\sigma]$	&	$4.5\sigma$ $[6.3\sigma]$ \\
\bottomrule
\end{tabular}
}

%% file: sections/Conclusions.tex

\section{Conclusions}
\label{sec:Conclusions}

In this work, we have presented an updated dispersive analysis of the pion VFF $F_\pi^V(s)$, building on top of previous work~\cite{Colangelo:2018mtw,Colangelo:2020lcg,Colangelo:2022prz,Stoffer:2023gba}. We have identified a connection between the largest systematic uncertainties and complex zeros in the VFF: they appear in the fits when using higher-order conformal polynomials to describe the inelastic effects and are connected to uncontrolled variations in the fit results. As such, their position randomly fluctuates with different orders of the conformal polynomial and different experimental data included in the fit. Given existing arguments in favor of the absence of zeros~\cite{Leutwyler:2002hm,Ananthanarayan:2011xt,RuizArriola:2024gwb}, we have analyzed the consequence of imposing that the VFF have no zeros. We observe that the fits stabilize in this case without any significant impact on the $p$-value, hence the current data do not favor zeros in the VFF. We have performed a comprehensive analysis of high-statistics $e^+e^-$ data based on the Omn\`es representation of the VFF, both with and without a constraint that imposes the absence of zeros. In addition, we have employed a hybrid phase-modulus dispersion relation that combines the knowledge about the elastic phase from solutions of the Roy equations with information on the modulus of the VFF above the inelastic threshold, relying on an explicit parametrization and BaBar data above $1.4\GeV$. We find that the two representations lead to compatible results.

Implementing the constraints of analyticity and unitarity, the dispersive analysis of the VFF demonstrates that the discrepancies in the two-pion contribution to the muon anomalous magnetic moment $a_\mu^{\pi\pi}$ are larger than what one would deduce by directly integrating the cross section in the limited energy range measured by each experiment. In particular, the tensions between CMD-3 and all the other experiments propagate outside the $\rho$-resonance region: the dispersive constraints correlate them to tensions at very low energies, e.g., in very-long-distance Euclidean windows. Tensions are present also in the pion charge radius, which thereby becomes an interesting observable for independent checks with improved lattice-QCD computations. In addition, the hybrid representation reveals that there are tensions between CMD-3 and BaBar data even above $1\GeV$, reflected by a much lower $p$-value of the hybrid fit to CMD-3 than the low-energy Omn\`es representation.

While most of our general conclusions hold in any dispersive representation, the tensions are certainly enhanced if the dominant systematic uncertainties are suppressed by imposing the absence of zeros. Comparing CMD-3 with KLOE, the discrepancies reach the level of $9\sigma$ in the contribution to $a_\mu$, and almost $5\sigma$ in the pion charge radius. This conclusion is reached when treating systematic uncertainties in the CMD-3 data as fully correlated, making use of established unbiased routines. We studied simple scenarios of de-correlation of the systematic uncertainties, invariably finding a further enhancement of the discrepancies.

Despite intense past and ongoing scrutiny, the origin of the discrepancies remains elusive. Understanding the reason for the current puzzling situation is indispensable, in order to obtain a consolidated data-driven prediction for the HVP contribution to $a_\mu$.

%% file: sections/Hybrid.tex

\section{Rearrangement of the hybrid representation}
\label{app:Hybrid}

Omitting the sum-rule corrections and the $\omega\phi$ factors, the hybrid phase-modulus representation can be written as
\begin{equation}\label{eq:Fhybrid2}
	F_\pi^V(s)=\Omega_1^1(s)\times \mathcal{DR}\left[|\tilde{F}/\Omega_1^1|;s_\inel\right](s) \, ,
\end{equation}
where $\tilde{F}$ can be any function whose modulus agrees with the modulus of the VFF on the inelastic branch cut. From the definition of the modulus dispersion relation~\eqref{eq:defDR}, we deduce immediately the identity
\begin{equation}
	\mathcal{DR}\left[|\tilde{F}/\Omega_1^1|;s_\inel\right](s) = \frac{\mathcal{DR}\bigl[ |\tilde{F}|;s_\inel \bigr](s)}{\mathcal{DR}\bigl[ |\Omega_1^1|;s_\inel \bigr](s)} \, ,
\end{equation}
and the factors of the product~\eqref{eq:Fhybrid2} can be rearranged as
\begin{equation}
	F_\pi^V(s) = \bar{\omega}_1^1(s)\times\mathcal{DR}\left[ |\tilde{F}|;s_\inel\right](s) \, ,
\end{equation}
with
\begin{equation}
	\bar{\omega}_1^1(s) = \frac{\Omega_1^1(s)}{\mathcal{DR}\left[ |\Omega_1^1|;s_\inel\right](s)} \, .
\end{equation}

One can write $\bar{\omega}^1_1$ in terms of the elastic phase shift $\delta_1^1$ in a simpler way, by using a once-subtracted dispersion relation on the auxiliary function
\begin{equation}
    f(s)=\dfrac{\log\bar{\omega}_1^1(s)-\left(\frac{s-\sthr}{s_\inel-\sthr}\right)^{3/2}\log\bar{\omega}_1^1(s_\inel)}{(s_\inel-s)^{3/2}},
\end{equation}
that automatically implements the $P$-wave behavior at the two-pion threshold $\sthr$ and the inelastic threshold $s_\inel$, similarly to Eq.~\eqref{eq:defDR}. The imaginary part of this function is given by
\begin{equation}
    \left\{
    \begin{array}{ll}
        \text{Im}\,f(s)=0 & \text{if } s<\sthr \, , \\
        \text{Im}\,f(s)=\dfrac{\delta_1^1(s)-\left(\frac{s-\sthr}{s_\inel-\sthr}\right)^{3/2}\delta_1^1(s_\inel)}{(s_\inel-s)^{3/2}} & \text{if }\sthr<s<s_\inel \, , \\
        \text{Im}\,f(s)=0 & \text{if }s_\inel<s \, .
    \end{array}
    \right.
\end{equation}
Therefore, the dispersion relation is written only in terms of an integral on the finite interval $[\sthr;s_\inel]$, leading to Eq.~\eqref{omegabar11}.

%% file: sections/Radius.tex

\section{Charge radius in the dispersive representations}
\label{app:Radius}

Since $F_\pi^V(s)$ is written as a product of functions that are all normalized to $1$ at $s=0$, we can decompose its derivative at $s=0$ as the sum of the derivatives of all the factors at $s=0$. For the Omn\`es representation valid at low energies~\eqref{eq:VFFOmnesRepresentation}, we write
\begin{equation}
	\< r_\pi^2\> = \< r_\pi^2\>_\text{el} + \< r_\pi^2\>_\omega + \< r_\pi^2\>_\text{inel} \, ,
\end{equation}
whereas the hybrid phase-modulus representation~\eqref{eq:HybridRepresentation} is expressed as
\begin{equation}
    \< r_\pi^2\>=\< r_\pi^2\>_{\overline{\text{el}}}+\< r_\pi^2\>_\omega+\< r_\pi^2\>_\phi+\< r_\pi^2\>_{|\tilde F|}+\< r_\pi^2\>_\infty \, ,
\end{equation}
where the different contributions are specified in the following. The elastic contribution in the Omn\`es representation reads
\begin{equation}
	\< r_\pi^2\>_{\text{el}} = 6\left.\frac{d\Omega_1^1(s)}{ds}\right|_{s=0} = \frac{6}{\pi}\int_{4M_\pi^2}^\infty ds\frac{\delta_1^1(s)}{s^2} \, .
\end{equation}
The contribution of a dispersively improved narrow resonance is
\begin{align}
	\< r_\pi^2\>_V = 6\left.\frac{dG_V(s)}{ds}\right|_{s=0}
		&= \frac{6}{\pi}\int_{9M_\pi^2}^\infty ds\frac{\text{Re}\,\epsilon_V}{s^2}\text{Im}\left[\frac{s}{\left(M_V-\frac{i}{2}\Gamma_V\right)^2-s}\right]\left(\frac{1-\frac{9M_\pi^2}{s}}{1-\frac{9M_\pi^2}{M_V^2}}\right)^4 \nn
	        &\quad +\frac{6}{\pi}\int_{M_{\pi^0}^2}^\infty ds\frac{\text{Im}\,\epsilon_V}{s^2}\text{Re}\left[\frac{s}{\left(M_V-\frac{i}{2}\Gamma_V\right)^2-s}\right]\left(\frac{1-\frac{M_{\pi^0}^2}{s}}{1-\frac{M_{\pi^0}^2}{M_V^2}}\right)^3 \, ,
\end{align}
which in principle can also be integrated analytically. The inelastic contribution from the conformal polynomial is given by
\begin{equation}
	\< r_\pi^2\>_\inel = 6\left.\frac{d G_\inel(s)}{ds}\right|_{s=0} = 6\frac{\sqrt{s_\inel-s_c}}{\sqrt{s_\inel}(\sqrt{s_\inel}+\sqrt{s_\inel-s_c})^2}P_N'(z_0) \, .
\end{equation}
The elastic contribution in the hybrid representation reads
\begin{equation}
	\< r_\pi^2\>_{\overline{\text{el}}} = -\frac{9}{s_\inel} \left(\frac{\sthr}{s_\inel - \sthr}\right)^{1/2} \delta_1^1(s_\inel) + \frac{6s_\inel^{3/2}}{\pi}\int_{\sthr}^{s_\inel}ds\frac{\delta_1^1(s)-\left(\frac{s-\sthr}{s_\inel-\sthr}\right)^{3/2}\delta_1^1(s_\inel)}{s^2(s_\inel-s)^{3/2}} \, ,
\end{equation}
whereas the inelastic contribution from the modulus dispersion integral is
\begin{equation}
	\< r_\pi^2\>_{|\tilde F|} = \frac{9}{s_\inel}\log|\tilde{F}(s_\inel)| - \frac{6s_\inel^{3/2}}{\pi}\int_{s_\inel}^\infty ds\frac{\log|\tilde{F}(s)/\tilde{F}(s_\inel)|}{s^2(s-s_\inel)^{3/2}} \, .
\end{equation}
The sum-rule contribution coming from the correction factor $G_\infty(s)$ is given by
\begin{equation}
	\< r_\pi^2\>_\infty = \frac{9(4a+5b)}{4s_\inel} \, .
\end{equation}
Similarly to Eq.~\eqref{eq:HybridRepresentation2}, in the numerical implementation it can be advantageous to split the sum-rule contribution into a part due to $1/|\Omega_1^1|$ and another piece depending on $|\tilde F|$, which can be computed together with $\< r_\pi^2\>_{|\tilde F|}$.

%% file: Paper.bbl
\providecommand{\href}[2]{#2}\begingroup\raggedright\begin{thebibliography}{100}

\bibitem{Muong-2:2021ojo}
B.~Abi {\em et~al.} [Muon g-2 Collaboration], Phys. Rev. Lett. {\bfseries 126},
  141801 (2021),
  [\href{https://arxiv.org/abs/2104.03281}{{arXiv:2104.03281~[hep-ex]}}].

\bibitem{Muong-2:2023cdq}
D.~P. Aguillard {\em et~al.} [Muon g-2 Collaboration], Phys. Rev. Lett.
  {\bfseries 131}, 161802 (2023),
  [\href{https://arxiv.org/abs/2308.06230}{{arXiv:2308.06230~[hep-ex]}}].

\bibitem{Muong-2:2024hpx}
D.~P. Aguillard {\em et~al.} [Muon g-2 Collaboration], Phys. Rev. D {\bfseries
  110}, 032009 (2024),
  [\href{https://arxiv.org/abs/2402.15410}{{arXiv:2402.15410~[hep-ex]}}].

\bibitem{Muong-2:2006rrc}
G.~W. Bennett {\em et~al.} [Muon g-2 Collaboration], Phys. Rev. D {\bfseries
  73}, 072003 (2006),
  [\href{https://arxiv.org/abs/hep-ex/0602035}{{arXiv:hep-ex/0602035}}].

\bibitem{Aoyama:2020ynm}
T.~Aoyama {\em et~al.}, Phys. Rept. {\bfseries 887}, 1 (2020),
  [\href{https://arxiv.org/abs/2006.04822}{{arXiv:2006.04822~[hep-ph]}}].

\bibitem{Colangelo:2022jxc}
G.~Colangelo {\em et~al.},
  \href{https://arxiv.org/abs/2203.15810}{{arXiv:2203.15810~[hep-ph]}}.

\bibitem{Blum:2019ugy}
T.~Blum, N.~Christ, M.~Hayakawa, T.~Izubuchi, L.~Jin, C.~Jung, and C.~Lehner,
  Phys. Rev. Lett. {\bfseries 124}, 132002 (2020),
  [\href{https://arxiv.org/abs/1911.08123}{{arXiv:1911.08123~[hep-lat]}}].

\bibitem{Chao:2021tvp}
E.-H. Chao, R.~J. Hudspith, A.~G\'erardin, J.~R. Green, H.~B. Meyer, and
  K.~Ottnad, Eur. Phys. J. C {\bfseries 81}, 651 (2021),
  [\href{https://arxiv.org/abs/2104.02632}{{arXiv:2104.02632~[hep-lat]}}].

\bibitem{Blum:2023vlm}
T.~Blum, N.~Christ, M.~Hayakawa, T.~Izubuchi, L.~Jin, C.~Jung, C.~Lehner, and
  C.~Tu, \href{https://arxiv.org/abs/2304.04423}{{arXiv:2304.04423~[hep-lat]}}.

\bibitem{Fodor:2024jyn}
Z.~Fodor, A.~G\'erardin, L.~Lellouch, K.~K. Szabo, B.~C. Toth, and
  C.~Zimmermann,
  \href{https://arxiv.org/abs/2411.11719}{{arXiv:2411.11719~[hep-lat]}}.

\bibitem{Melnikov:2003xd}
K.~Melnikov and A.~Vainshtein, Phys. Rev. D {\bfseries 70}, 113006 (2004),
[\href{https://arxiv.org/abs/hep-ph/0312226}{{arXiv:hep-ph/0312226}}].

\bibitem{Pauk:2014rta}
V.~Pauk and M.~Vanderhaeghen, Eur. Phys. J. C {\bfseries 74}, 3008 (2014),
  [\href{https://arxiv.org/abs/1401.0832}{{arXiv:1401.0832~[hep-ph]}}].

\bibitem{Colangelo:2014dfa}
G.~Colangelo, M.~Hoferichter, M.~Procura, and P.~Stoffer, JHEP {\bfseries 09},
  091 (2014),
  [\href{https://arxiv.org/abs/1402.7081}{{arXiv:1402.7081~[hep-ph]}}].

\bibitem{Colangelo:2014qya}
G.~Colangelo, M.~Hoferichter, A.~Nyffeler, M.~Passera, and P.~Stoffer, Phys.
  Lett. B {\bfseries 735}, 90 (2014),
  [\href{https://arxiv.org/abs/1403.7512}{{arXiv:1403.7512~[hep-ph]}}].

\bibitem{Colangelo:2014pva}
G.~Colangelo, M.~Hoferichter, B.~Kubis, M.~Procura, and P.~Stoffer, Phys. Lett.
  B {\bfseries 738}, 6 (2014),
  [\href{https://arxiv.org/abs/1408.2517}{{arXiv:1408.2517~[hep-ph]}}].

\bibitem{Colangelo:2015ama}
G.~Colangelo, M.~Hoferichter, M.~Procura, and P.~Stoffer, JHEP {\bfseries 09},
  074 (2015),
[\href{https://arxiv.org/abs/1506.01386}{{arXiv:1506.01386~[hep-ph]}}].

\bibitem{Danilkin:2016hnh}
I.~Danilkin and M.~Vanderhaeghen, Phys. Rev. D {\bfseries 95}, 014019 (2017),
  [\href{https://arxiv.org/abs/1611.04646}{{arXiv:1611.04646~[hep-ph]}}].

\bibitem{Masjuan:2017tvw}
P.~Masjuan and P.~S\'anchez-Puertas, Phys. Rev. D {\bfseries 95}, 054026
  (2017),
  [\href{https://arxiv.org/abs/1701.05829}{{arXiv:1701.05829~[hep-ph]}}].

\bibitem{Colangelo:2017qdm}
G.~Colangelo, M.~Hoferichter, M.~Procura, and P.~Stoffer, Phys. Rev. Lett.
  {\bfseries 118}, 232001 (2017),
  [\href{https://arxiv.org/abs/1701.06554}{{arXiv:1701.06554~[hep-ph]}}].

\bibitem{Colangelo:2017fiz}
G.~Colangelo, M.~Hoferichter, M.~Procura, and P.~Stoffer, JHEP {\bfseries 04},
  161 (2017),
[\href{https://arxiv.org/abs/1702.07347}{{arXiv:1702.07347~[hep-ph]}}].

\bibitem{Jegerlehner:2017gek}
F.~Jegerlehner, \href{http://dx.doi.org/10.1007/978-3-319-63577-4}{{\em {The
  Anomalous Magnetic Moment of the Muon}}}, vol.~274.
\newblock Springer, Cham, 2017.

\bibitem{Knecht:2018sci}
M.~Knecht, S.~Narison, A.~Rabemananjara, and D.~Rabetiarivony, Phys. Lett. B
  {\bfseries 787}, 111 (2018),
  [\href{https://arxiv.org/abs/1808.03848}{{arXiv:1808.03848~[hep-ph]}}].

\bibitem{Hoferichter:2018kwz}
M.~Hoferichter, B.-L. Hoid, B.~Kubis, S.~Leupold, and S.~P. Schneider, JHEP
  {\bfseries 10}, 141 (2018),
[\href{https://arxiv.org/abs/1808.04823}{{arXiv:1808.04823~[hep-ph]}}].

\bibitem{Gerardin:2019vio}
A.~G\'erardin, H.~B. Meyer, and A.~Nyffeler, Phys. Rev. D {\bfseries 100},
  034520 (2019),
  [\href{https://arxiv.org/abs/1903.09471}{{arXiv:1903.09471~[hep-lat]}}].

\bibitem{Hoferichter:2019nlq}
M.~Hoferichter and P.~Stoffer, JHEP {\bfseries 07}, 073 (2019),
  [\href{https://arxiv.org/abs/1905.13198}{{arXiv:1905.13198~[hep-ph]}}].

\bibitem{Bijnens:2019ghy}
J.~Bijnens, N.~Hermansson-Truedsson, and A.~Rodr\'\i{}guez-S\'anchez, Phys.
  Lett. B {\bfseries 798}, 134994 (2019),
  [\href{https://arxiv.org/abs/1908.03331}{{arXiv:1908.03331~[hep-ph]}}].

\bibitem{Roig:2019reh}
P.~Roig and P.~S\'anchez-Puertas, Phys. Rev. D {\bfseries 101}, 074019 (2020),
  [\href{https://arxiv.org/abs/1910.02881}{{arXiv:1910.02881~[hep-ph]}}].

\bibitem{Eichmann:2019bqf}
G.~Eichmann, C.~S. Fischer, and R.~Williams, Phys. Rev. D {\bfseries 101},
  054015 (2020),
  [\href{https://arxiv.org/abs/1910.06795}{{arXiv:1910.06795~[hep-ph]}}].

\bibitem{Colangelo:2019lpu}
G.~Colangelo, F.~Hagelstein, M.~Hoferichter, L.~Laub, and P.~Stoffer, Phys.
  Rev. D {\bfseries 101}, 051501 (2020),
[\href{https://arxiv.org/abs/1910.11881}{{arXiv:1910.11881~[hep-ph]}}].

\bibitem{Colangelo:2019uex}
G.~Colangelo, F.~Hagelstein, M.~Hoferichter, L.~Laub, and P.~Stoffer, JHEP
  {\bfseries 03}, 101 (2020),
  [\href{https://arxiv.org/abs/1910.13432}{{arXiv:1910.13432~[hep-ph]}}].

\bibitem{Ludtke:2020moa}
J.~L\"udtke and M.~Procura, Eur. Phys. J. C {\bfseries 80}, 1108 (2020),
  [\href{https://arxiv.org/abs/2006.00007}{{arXiv:2006.00007~[hep-ph]}}].

\bibitem{Bijnens:2020xnl}
J.~Bijnens, N.~Hermansson-Truedsson, L.~Laub, and A.~Rodr\'\i{}guez-S\'anchez,
  JHEP {\bfseries 10}, 203 (2020),
  [\href{https://arxiv.org/abs/2008.13487}{{arXiv:2008.13487~[hep-ph]}}].

\bibitem{Bijnens:2021jqo}
J.~Bijnens, N.~Hermansson-Truedsson, L.~Laub, and A.~Rodr\'\i{}guez-S\'anchez,
  JHEP {\bfseries 04}, 240 (2021),
  [\href{https://arxiv.org/abs/2101.09169}{{arXiv:2101.09169~[hep-ph]}}].

\bibitem{Danilkin:2021icn}
I.~Danilkin, M.~Hoferichter, and P.~Stoffer, Phys. Lett. B {\bfseries 820},
  136502 (2021),
  [\href{https://arxiv.org/abs/2105.01666}{{arXiv:2105.01666~[hep-ph]}}].

\bibitem{Colangelo:2021nkr}
G.~Colangelo, F.~Hagelstein, M.~Hoferichter, L.~Laub, and P.~Stoffer, Eur.
  Phys. J. C {\bfseries 81}, 702 (2021),
  [\href{https://arxiv.org/abs/2106.13222}{{arXiv:2106.13222~[hep-ph]}}].

\bibitem{Stamen:2022uqh}
D.~Stamen, D.~Hariharan, M.~Hoferichter, B.~Kubis, and P.~Stoffer, Eur. Phys.
  J. C {\bfseries 82}, 432 (2022),
  [\href{https://arxiv.org/abs/2202.11106}{{arXiv:2202.11106~[hep-ph]}}].

\bibitem{Bijnens:2022itw}
J.~Bijnens, N.~Hermansson-Truedsson, and A.~Rodr\'\i{}guez-S\'anchez, JHEP
  {\bfseries 02}, 167 (2023),
  [\href{https://arxiv.org/abs/2211.17183}{{arXiv:2211.17183~[hep-ph]}}].

\bibitem{Ludtke:2023hvz}
J.~L\"udtke, M.~Procura, and P.~Stoffer, JHEP {\bfseries 04}, 125 (2023),
  [\href{https://arxiv.org/abs/2302.12264}{{arXiv:2302.12264~[hep-ph]}}].

\bibitem{Stoffer:2023gba}
P.~Stoffer, G.~Colangelo, and M.~Hoferichter, JINST {\bfseries 18}, C10021
  (2023),
  [\href{https://arxiv.org/abs/2308.04217}{{arXiv:2308.04217~[hep-ph]}}].

\bibitem{Hoferichter:2024fsj}
M.~Hoferichter, P.~Stoffer, and M.~Zillinger, JHEP {\bfseries 04}, 092 (2024),
  [\href{https://arxiv.org/abs/2402.14060}{{arXiv:2402.14060~[hep-ph]}}].

\bibitem{Ludtke:2024ase}
J.~L\"udtke, M.~Procura, and P.~Stoffer,
  \href{https://arxiv.org/abs/2410.11946}{{arXiv:2410.11946~[hep-ph]}}.

\bibitem{Holz:2024lom}
S.~Holz, M.~Hoferichter, B.-L. Hoid, and B.~Kubis,
  \href{https://arxiv.org/abs/2411.08098}{{arXiv:2411.08098~[hep-ph]}}.

\bibitem{Holz:2024diw}
S.~Holz, M.~Hoferichter, B.-L. Hoid, and B.~Kubis,
  \href{https://arxiv.org/abs/2412.16281}{{arXiv:2412.16281~[hep-ph]}}.

\bibitem{Hoferichter:2024bae}
M.~Hoferichter, P.~Stoffer, and M.~Zillinger,
  \href{https://arxiv.org/abs/2412.00178}{{arXiv:2412.00178~[hep-ph]}}.

\bibitem{Hoferichter:2024vbu}
M.~Hoferichter, P.~Stoffer, and M.~Zillinger,
  \href{https://arxiv.org/abs/2412.00190}{{arXiv:2412.00190~[hep-ph]}}.

\bibitem{Borsanyi:2020mff}
S.~Borsanyi {\em et~al.}, Nature {\bfseries 593}, 51 (2021),
  [\href{https://arxiv.org/abs/2002.12347}{{arXiv:2002.12347~[hep-lat]}}].

\bibitem{Davier:2017zfy}
M.~Davier, A.~Hoecker, B.~Malaescu, and Z.~Zhang, Eur. Phys. J. C {\bfseries
  77}, 827 (2017),
  [\href{https://arxiv.org/abs/1706.09436}{{arXiv:1706.09436~[hep-ph]}}].

\bibitem{Keshavarzi:2018mgv}
A.~Keshavarzi, D.~Nomura, and T.~Teubner, Phys. Rev. D {\bfseries 97}, 114025
  (2018),
  [\href{https://arxiv.org/abs/1802.02995}{{arXiv:1802.02995~[hep-ph]}}].

\bibitem{Colangelo:2018mtw}
G.~Colangelo, M.~Hoferichter, and P.~Stoffer, JHEP {\bfseries 02}, 006 (2019),
  [\href{https://arxiv.org/abs/1810.00007}{{arXiv:1810.00007~[hep-ph]}}].

\bibitem{Hoferichter:2019mqg}
M.~Hoferichter, B.-L. Hoid, and B.~Kubis, JHEP {\bfseries 08}, 137 (2019),
  [\href{https://arxiv.org/abs/1907.01556}{{arXiv:1907.01556~[hep-ph]}}].

\bibitem{Davier:2019can}
M.~Davier, A.~Hoecker, B.~Malaescu, and Z.~Zhang, Eur. Phys. J. C {\bfseries
  80}, 241 (2020),
  [\href{https://arxiv.org/abs/1908.00921}{{arXiv:1908.00921~[hep-ph]}}],
  [Erratum: Eur.~Phys.~J.~C {\bf 80}, 410 (2020)].

\bibitem{Keshavarzi:2019abf}
A.~Keshavarzi, D.~Nomura, and T.~Teubner, Phys. Rev. D {\bfseries 101}, 014029
  (2020),
  [\href{https://arxiv.org/abs/1911.00367}{{arXiv:1911.00367~[hep-ph]}}].

\bibitem{Hoid:2020xjs}
B.-L. Hoid, M.~Hoferichter, and B.~Kubis, Eur. Phys. J. C {\bfseries 80}, 988
  (2020),
  [\href{https://arxiv.org/abs/2007.12696}{{arXiv:2007.12696~[hep-ph]}}].

\bibitem{Colangelo:2021moe}
G.~Colangelo, M.~Hoferichter, B.~Kubis, M.~Niehus, and J.~Ruiz~de Elvira, Phys.
  Lett. B {\bfseries 825}, 136852 (2022),
  [\href{https://arxiv.org/abs/2110.05493}{{arXiv:2110.05493~[hep-ph]}}].

\bibitem{Colangelo:2022prz}
G.~Colangelo, M.~Hoferichter, B.~Kubis, and P.~Stoffer, JHEP {\bfseries 10},
  032 (2022),
  [\href{https://arxiv.org/abs/2208.08993}{{arXiv:2208.08993~[hep-ph]}}].

\bibitem{Hoferichter:2023sli}
M.~Hoferichter, G.~Colangelo, B.-L. Hoid, B.~Kubis, J.~Ruiz~de Elvira,
  D.~Schuh, D.~Stamen, and P.~Stoffer, Phys. Rev. Lett. {\bfseries 131}, 161905
  (2023),
  [\href{https://arxiv.org/abs/2307.02532}{{arXiv:2307.02532~[hep-ph]}}].

\bibitem{Ce:2022kxy}
M.~C\`e {\em et~al.}, Phys. Rev. D {\bfseries 106}, 114502 (2022),
  [\href{https://arxiv.org/abs/2206.06582}{{arXiv:2206.06582~[hep-lat]}}].

\bibitem{ExtendedTwistedMass:2022jpw}
C.~Alexandrou {\em et~al.} [Extended Twisted Mass Collaboration], Phys. Rev. D
  {\bfseries 107}, 074506 (2023),
  [\href{https://arxiv.org/abs/2206.15084}{{arXiv:2206.15084~[hep-lat]}}].

\bibitem{FermilabLatticeHPQCD:2023jof}
A.~Bazavov {\em et~al.} [Fermilab Lattice, HPQCD,, MILC Collaboration], Phys.
  Rev. D {\bfseries 107}, 114514 (2023),
  [\href{https://arxiv.org/abs/2301.08274}{{arXiv:2301.08274~[hep-lat]}}].

\bibitem{RBC:2023pvn}
T.~Blum {\em et~al.} [RBC, UKQCD Collaboration], Phys. Rev. D {\bfseries 108},
  054507 (2023),
  [\href{https://arxiv.org/abs/2301.08696}{{arXiv:2301.08696~[hep-lat]}}].

\bibitem{Boccaletti:2024guq}
A.~Boccaletti {\em et~al.},
  \href{https://arxiv.org/abs/2407.10913}{{arXiv:2407.10913~[hep-lat]}}.

\bibitem{RBC:2024fic}
T.~Blum {\em et~al.} [RBC, UKQCD Collaboration],
  \href{https://arxiv.org/abs/2410.20590}{{arXiv:2410.20590~[hep-lat]}}.

\bibitem{Djukanovic:2024cmq}
D.~Djukanovic, G.~von Hippel, S.~Kuberski, H.~B. Meyer, N.~Miller, K.~Ottnad,
  J.~Parrino, A.~Risch, and H.~Wittig,
  \href{https://arxiv.org/abs/2411.07969}{{arXiv:2411.07969~[hep-lat]}}.

\bibitem{Bazavov:2024eou}
A.~Bazavov {\em et~al.},
  \href{https://arxiv.org/abs/2412.18491}{{arXiv:2412.18491~[hep-lat]}}.

\bibitem{CMD-3:2023rfe}
F.~V. Ignatov {\em et~al.} [CMD-3 Collaboration], Phys. Rev. Lett. {\bfseries
  132}, 231903 (2024),
  [\href{https://arxiv.org/abs/2309.12910}{{arXiv:2309.12910~[hep-ex]}}].

\bibitem{CMD-3:2023alj}
F.~V. Ignatov {\em et~al.} [CMD-3 Collaboration], Phys. Rev. D {\bfseries 109},
  112002 (2024),
  [\href{https://arxiv.org/abs/2302.08834}{{arXiv:2302.08834~[hep-ex]}}].

\bibitem{Crivellin:2020zul}
A.~Crivellin, M.~Hoferichter, C.~A. Manzari, and M.~Montull, Phys. Rev. Lett.
  {\bfseries 125}, 091801 (2020),
  [\href{https://arxiv.org/abs/2003.04886}{{arXiv:2003.04886~[hep-ph]}}].

\bibitem{Keshavarzi:2020bfy}
A.~Keshavarzi, W.~J. Marciano, M.~Passera, and A.~Sirlin, Phys. Rev. D
  {\bfseries 102}, 033002 (2020),
  [\href{https://arxiv.org/abs/2006.12666}{{arXiv:2006.12666~[hep-ph]}}].

\bibitem{Malaescu:2020zuc}
B.~Malaescu and M.~Schott, Eur. Phys. J. C {\bfseries 81}, 46 (2021),
  [\href{https://arxiv.org/abs/2008.08107}{{arXiv:2008.08107~[hep-ph]}}].

\bibitem{Colangelo:2020lcg}
G.~Colangelo, M.~Hoferichter, and P.~Stoffer, Phys. Lett. B {\bfseries 814},
  136073 (2021),
  [\href{https://arxiv.org/abs/2010.07943}{{arXiv:2010.07943~[hep-ph]}}].

\bibitem{Campanario:2019mjh}
F.~Campanario, H.~Czy\.z, J.~Gluza, T.~Jeli\'nski, G.~Rodrigo, S.~Tracz, and
  D.~Zhuridov, Phys. Rev. D {\bfseries 100}, 076004 (2019),
  [\href{https://arxiv.org/abs/1903.10197}{{arXiv:1903.10197~[hep-ph]}}].

\bibitem{Monnard:2021pvm}
J.~Monnard, PhD thesis, Universit\"at Bern  (2021).
\newblock \url{https://boristheses.unibe.ch/2825/}.

\bibitem{Ignatov:2022iou}
F.~Ignatov and R.~N. Lee, Phys. Lett. B {\bfseries 833}, 137283 (2022),
  [\href{https://arxiv.org/abs/2204.12235}{{arXiv:2204.12235~[hep-ph]}}].

\bibitem{Colangelo:2022lzg}
G.~Colangelo, M.~Hoferichter, J.~Monnard, and J.~Ruiz~de Elvira, JHEP
  {\bfseries 08}, 295 (2022),
  [\href{https://arxiv.org/abs/2207.03495}{{arXiv:2207.03495~[hep-ph]}}],
  [Erratum: JHEP {\bf 09}, 177 (2024)].

\bibitem{BaBar:2023xiy}
J.~P. Lees {\em et~al.} [BaBar Collaboration], Phys. Rev. D {\bfseries 108},
  L111103 (2023),
  [\href{https://arxiv.org/abs/2308.05233}{{arXiv:2308.05233~[hep-ex]}}].

\bibitem{Budassi:2024whw}
E.~Budassi, C.~M. Carloni~Calame, M.~Ghilardi, A.~Gurgone, G.~Montagna,
  M.~Moretti, O.~Nicrosini, F.~Piccinini, and F.~P. Ucci,
  \href{https://arxiv.org/abs/2409.03469}{{arXiv:2409.03469~[hep-ph]}}.

\bibitem{Abbiendi:2022liz}
G.~Abbiendi {\em et~al.},
  \href{https://arxiv.org/abs/2201.12102}{{arXiv:2201.12102~[hep-ph]}}.

\bibitem{Aliberti:2024fpq}
R.~Aliberti {\em et~al.},
  \href{https://arxiv.org/abs/2410.22882}{{arXiv:2410.22882~[hep-ph]}}.

\bibitem{Leutwyler:2002hm}
H.~Leutwyler, in: {\em Continuous advances in QCD 2002,} {\bfseries \normalfont
  {eds. K.~A.~Olive, M.~A.~Shifman, and M.~B.~Voloshin, World Scientific,
  Singapore, pp. 23--40}},  (2003),
[\href{https://arxiv.org/abs/hep-ph/0212324}{{arXiv:hep-ph/0212324}}].

\bibitem{Colangelo:2003yw}
G.~Colangelo, Nucl. Phys. B Proc. Suppl. {\bfseries 131}, 185 (2004),
  [\href{https://arxiv.org/abs/hep-ph/0312017}{{arXiv:hep-ph/0312017}}].

\bibitem{Ananthanarayan:2011xt}
B.~Ananthanarayan, I.~Caprini, and I.~S. Imsong, Phys. Rev. D {\bfseries 83},
  096002 (2011),
  [\href{https://arxiv.org/abs/1102.3299}{{arXiv:1102.3299~[hep-ph]}}].

\bibitem{Hanhart:2012wi}
C.~Hanhart, Phys. Lett. B {\bfseries 715}, 170 (2012),
  [\href{https://arxiv.org/abs/1203.6839}{{arXiv:1203.6839~[hep-ph]}}].

\bibitem{Ananthanarayan:2012tt}
B.~Ananthanarayan, I.~Caprini, D.~Das, and I.~S. Imsong, Eur. Phys. J. C
  {\bfseries 72}, 2192 (2012),
  [\href{https://arxiv.org/abs/1209.0379}{{arXiv:1209.0379~[hep-ph]}}].

\bibitem{Ananthanarayan:2013zua}
B.~Ananthanarayan, I.~Caprini, D.~Das, and I.~Sentitemsu~Imsong, Phys. Rev. D
  {\bfseries 89}, 036007 (2014),
  [\href{https://arxiv.org/abs/1312.5849}{{arXiv:1312.5849~[hep-ph]}}].

\bibitem{Ananthanarayan:2016mns}
B.~Ananthanarayan, I.~Caprini, D.~Das, and I.~Sentitemsu~Imsong, Phys. Rev. D
  {\bfseries 93}, 116007 (2016),
  [\href{https://arxiv.org/abs/1605.00202}{{arXiv:1605.00202~[hep-ph]}}].

\bibitem{Ananthanarayan:2017efc}
B.~Ananthanarayan, I.~Caprini, and D.~Das, Phys. Rev. Lett. {\bfseries 119},
  132002 (2017),
  [\href{https://arxiv.org/abs/1706.04020}{{arXiv:1706.04020~[hep-ph]}}].

\bibitem{Ananthanarayan:2018nyx}
B.~Ananthanarayan, I.~Caprini, and D.~Das, Phys. Rev. D {\bfseries 98}, 114015
  (2018),
  [\href{https://arxiv.org/abs/1810.09265}{{arXiv:1810.09265~[hep-ph]}}].

\bibitem{Ananthanarayan:2020vum}
B.~Ananthanarayan, I.~Caprini, and D.~Das, Phys. Rev. D {\bfseries 102}, 096003
  (2020),
  [\href{https://arxiv.org/abs/2008.00669}{{arXiv:2008.00669~[hep-ph]}}].

\bibitem{Simula:2023ujs}
S.~Simula and L.~Vittorio, Phys. Rev. D {\bfseries 108}, 094013 (2023),
  [\href{https://arxiv.org/abs/2309.02135}{{arXiv:2309.02135~[hep-ph]}}].

\bibitem{RuizArriola:2024gwb}
E.~Ruiz~Arriola and P.~S\'anchez-Puertas, Phys. Rev. D {\bfseries 110}, 054003
  (2024),
  [\href{https://arxiv.org/abs/2403.07121}{{arXiv:2403.07121~[hep-ph]}}].

\bibitem{Muskhelishvili:1953}
N.~I. Muskhelishvili, {\em {Singular integral equations}}.
\newblock Wolters-Noordhoff Publishing, Groningen, 1953.

\bibitem{Omnes:1958hv}
R.~Omn\`es, Nuovo Cim. {\bfseries 8}, 316 (1958).

\bibitem{Roy:1971tc}
S.~M. Roy, Phys. Lett. B {\bfseries 36}, 353 (1971).

\bibitem{Ananthanarayan:2000ht}
B.~Ananthanarayan, G.~Colangelo, J.~Gasser, and H.~Leutwyler, Phys. Rept.
  {\bfseries 353}, 207 (2001),
  [\href{https://arxiv.org/abs/hep-ph/0005297}{{arXiv:hep-ph/0005297}}].

\bibitem{Garcia-Martin:2011iqs}
R.~Garc\'ia-Mart\'in, R.~Kami\'nski, J.~R. Pel\'aez, J.~Ruiz~de Elvira, and
  F.~J. Yndur\'ain, Phys. Rev. D {\bfseries 83}, 074004 (2011),
  [\href{https://arxiv.org/abs/1102.2183}{{arXiv:1102.2183~[hep-ph]}}].

\bibitem{Caprini:2011ky}
I.~Caprini, G.~Colangelo, and H.~Leutwyler, Eur. Phys. J. C {\bfseries 72},
  1860 (2012),
  [\href{https://arxiv.org/abs/1111.7160}{{arXiv:1111.7160~[hep-ph]}}].

\bibitem{Pelaez:2024uav}
J.~R. Pel\'aez, P.~Rab\'an, and J.~Ruiz~de Elvira,
  \href{https://arxiv.org/abs/2412.15327}{{arXiv:2412.15327~[hep-ph]}}.

\bibitem{Hoferichter:2023bjm}
M.~Hoferichter, B.-L. Hoid, B.~Kubis, and D.~Schuh, JHEP {\bfseries 08}, 208
  (2023),
  [\href{https://arxiv.org/abs/2307.02546}{{arXiv:2307.02546~[hep-ph]}}].

\bibitem{Eidelman:2003uh}
S.~Eidelman and L.~\L{}ukaszuk, Phys. Lett. B {\bfseries 582}, 27 (2004),
  [\href{https://arxiv.org/abs/hep-ph/0311366}{{arXiv:hep-ph/0311366}}].

\bibitem{Lukaszuk:1973jd}
L.~\L{}ukaszuk, Phys. Lett. B {\bfseries 47}, 51 (1973).

\bibitem{ZerosInPrep:2025}
M.~Hoferichter, T.~P. Leplumey, and P.~Stoffer, work in progress.

\bibitem{Pham:1975at}
T.~N. Pham and T.~N. Truong, Phys. Rev. D {\bfseries 14}, 185 (1976).

\bibitem{Mohapatra:1977ht}
J.~K. Mohapatra and J.~Maharana, Phys. Rev. D {\bfseries 16}, 907 (1977).

\bibitem{Chanturia:2022rcz}
G.~Chanturia, PoS {\bfseries Regio2021}, 030 (2022).

\bibitem{Heuser:2024biq}
L.~A. Heuser, G.~Chanturia, F.~K. Guo, C.~Hanhart, M.~Hoferichter, and
  B.~Kubis, Eur. Phys. J. C {\bfseries 84}, 599 (2024),
  [\href{https://arxiv.org/abs/2403.15539}{{arXiv:2403.15539~[hep-ph]}}].

\bibitem{BaBar:2009wpw}
B.~Aubert {\em et~al.} [BaBar Collaboration], Phys. Rev. Lett. {\bfseries 103},
  231801 (2009),
  [\href{https://arxiv.org/abs/0908.3589}{{arXiv:0908.3589~[hep-ex]}}].

\bibitem{BaBar:2012bdw}
J.~P. Lees {\em et~al.} [BaBar Collaboration], Phys. Rev. D {\bfseries 86},
  032013 (2012),
  [\href{https://arxiv.org/abs/1205.2228}{{arXiv:1205.2228~[hep-ex]}}].

\bibitem{Gounaris:1968mw}
G.~J. Gounaris and J.~J. Sakurai, Phys. Rev. Lett. {\bfseries 21}, 244 (1968).

\bibitem{Watson:1954uc}
K.~M. Watson, Phys. Rev. {\bfseries 95}, 228 (1954).

\bibitem{PhysRev.172.1645}
T.~N. Truong, R.~V. Mau, and P.~X. Yem, Phys. Rev. {\bfseries 172}, 1645
  (1968).

\bibitem{Geshkenbein:1998gu}
B.~V. Geshkenbein, Phys. Rev. D {\bfseries 61}, 033009 (2000),
  [\href{https://arxiv.org/abs/hep-ph/9806418}{{arXiv:hep-ph/9806418}}].

\bibitem{Achasov:2005rg}
M.~N. Achasov {\em et~al.}, J. Exp. Theor. Phys. {\bfseries 101}, 1053 (2005),
  [\href{https://arxiv.org/abs/hep-ex/0506076}{{arXiv:hep-ex/0506076}}].

\bibitem{Achasov:2006vp}
M.~N. Achasov {\em et~al.}, J. Exp. Theor. Phys. {\bfseries 103}, 380 (2006),
  [\href{https://arxiv.org/abs/hep-ex/0605013}{{arXiv:hep-ex/0605013}}].

\bibitem{SND:2020nwa}
M.~N. Achasov {\em et~al.} [SND Collaboration], JHEP {\bfseries 01}, 113
  (2021),
  [\href{https://arxiv.org/abs/2004.00263}{{arXiv:2004.00263~[hep-ex]}}].

\bibitem{CMD-2:2001ski}
R.~R. Akhmetshin {\em et~al.} [CMD-2 Collaboration], Phys. Lett. B {\bfseries
  527}, 161 (2002),
  [\href{https://arxiv.org/abs/hep-ex/0112031}{{arXiv:hep-ex/0112031}}].

\bibitem{CMD-2:2003gqi}
R.~R. Akhmetshin {\em et~al.} [CMD-2 Collaboration], Phys. Lett. B {\bfseries
  578}, 285 (2004),
  [\href{https://arxiv.org/abs/hep-ex/0308008}{{arXiv:hep-ex/0308008}}].

\bibitem{Aulchenko:2006dxz}
V.~M. Aul'chenko {\em et~al.}, JETP Lett. {\bfseries 84}, 413 (2006),
  [\href{https://arxiv.org/abs/hep-ex/0610016}{{arXiv:hep-ex/0610016}}].

\bibitem{CMD-2:2006gxt}
R.~R. Akhmetshin {\em et~al.} [CMD-2 Collaboration], Phys. Lett. B {\bfseries
  648}, 28 (2007),
  [\href{https://arxiv.org/abs/hep-ex/0610021}{{arXiv:hep-ex/0610021}}].

\bibitem{KLOE:2008fmq}
F.~Ambrosino {\em et~al.} [KLOE Collaboration], Phys. Lett. B {\bfseries 670},
  285 (2009),
  [\href{https://arxiv.org/abs/0809.3950}{{arXiv:0809.3950~[hep-ex]}}].

\bibitem{KLOE:2010qei}
F.~Ambrosino {\em et~al.} [KLOE Collaboration], Phys. Lett. B {\bfseries 700},
  102 (2011),
  [\href{https://arxiv.org/abs/1006.5313}{{arXiv:1006.5313~[hep-ex]}}].

\bibitem{KLOE:2012anl}
D.~Babusci {\em et~al.} [KLOE Collaboration], Phys. Lett. B {\bfseries 720},
  336 (2013),
  [\href{https://arxiv.org/abs/1212.4524}{{arXiv:1212.4524~[hep-ex]}}].

\bibitem{KLOE-2:2017fda}
A.~Anastasi {\em et~al.} [KLOE-2 Collaboration], JHEP {\bfseries 03}, 173
  (2018),
  [\href{https://arxiv.org/abs/1711.03085}{{arXiv:1711.03085~[hep-ex]}}].

\bibitem{BESIII:2015equ}
M.~Ablikim {\em et~al.} [BESIII Collaboration], Phys. Lett. B {\bfseries 753},
  629 (2016),
  [\href{https://arxiv.org/abs/1507.08188}{{arXiv:1507.08188~[hep-ex]}}],
  [Erratum: Phys.Lett.B 812, 135982 (2021)].

\bibitem{NA7:1986vav}
S.~R. Amendolia {\em et~al.} [NA7 Collaboration], Nucl. Phys. B {\bfseries
  277}, 168 (1986).

\bibitem{Davier:2023fpl}
M.~Davier, A.~Hoecker, A.-M. Lutz, B.~Malaescu, and Z.~Zhang, Eur. Phys. J. C
  {\bfseries 84}, 721 (2024),
  [\href{https://arxiv.org/abs/2312.02053}{{arXiv:2312.02053~[hep-ph]}}].

\bibitem{Colangelo:2022vok}
G.~Colangelo, A.~X. El-Khadra, M.~Hoferichter, A.~Keshavarzi, C.~Lehner,
  P.~Stoffer, and T.~Teubner, Phys. Lett. B {\bfseries 833}, 137313 (2022),
  [\href{https://arxiv.org/abs/2205.12963}{{arXiv:2205.12963~[hep-ph]}}].

\bibitem{RBC:2018dos}
T.~Blum, P.~A. Boyle, V.~G\"ulpers, T.~Izubuchi, L.~Jin, C.~Jung, A.~J\"uttner,
  C.~Lehner, A.~Portelli, and J.~T. Tsang [RBC, UKQCD Collaboration], Phys.
  Rev. Lett. {\bfseries 121}, 022003 (2018),
  [\href{https://arxiv.org/abs/1801.07224}{{arXiv:1801.07224~[hep-lat]}}].

\bibitem{Lehner:2020crt}
C.~Lehner and A.~S. Meyer, Phys. Rev. D {\bfseries 101}, 074515 (2020),
  [\href{https://arxiv.org/abs/2003.04177}{{arXiv:2003.04177~[hep-lat]}}].

\bibitem{Gonzalez-Solis:2019iod}
S.~Gonz\`alez-Sol\'\i{}s and P.~Roig, Eur. Phys. J. C {\bfseries 79}, 436
  (2019),
  [\href{https://arxiv.org/abs/1902.02273}{{arXiv:1902.02273~[hep-ph]}}].

\bibitem{ParticleDataGroup:2024cfk}
S.~Navas {\em et~al.} [Particle Data Group], Phys. Rev. D {\bfseries 110},
  030001 (2024).

\bibitem{Koponen:2015tkr}
J.~Koponen, F.~Bursa, C.~T.~H. Davies, R.~J. Dowdall, and G.~P. Lepage, Phys.
  Rev. D {\bfseries 93}, 054503 (2016),
  [\href{https://arxiv.org/abs/1511.07382}{{arXiv:1511.07382~[hep-lat]}}].

\bibitem{Feng:2019geu}
X.~Feng, Y.~Fu, and L.-C. Jin, Phys. Rev. D {\bfseries 101}, 051502 (2020),
  [\href{https://arxiv.org/abs/1911.04064}{{arXiv:1911.04064~[hep-lat]}}].

\bibitem{Wang:2020nbf}
G.~Wang, J.~Liang, T.~Draper, K.-F. Liu, and Y.-B. Yang [$\chi$QCD
  Collaboration], Phys. Rev. D {\bfseries 104}, 074502 (2021),
  [\href{https://arxiv.org/abs/2006.05431}{{arXiv:2006.05431~[hep-ph]}}].

\bibitem{Gao:2021xsm}
X.~Gao, N.~Karthik, S.~Mukherjee, P.~Petreczky, S.~Syritsyn, and Y.~Zhao, Phys.
  Rev. D {\bfseries 104}, 114515 (2021),
  [\href{https://arxiv.org/abs/2102.06047}{{arXiv:2102.06047~[hep-lat]}}].

\bibitem{DAgostini:1993arp}
G.~D'Agostini, Nucl. Instrum. Meth. A {\bfseries 346}, 306 (1994).

\bibitem{Ball:2009qv}
R.~D. Ball, L.~Del~Debbio, S.~Forte, A.~Guffanti, J.~I. Latorre, J.~Rojo, and
  M.~Ubiali [NNPDF Collaboration], JHEP {\bfseries 05}, 075 (2010),
  [\href{https://arxiv.org/abs/0912.2276}{{arXiv:0912.2276~[hep-ph]}}].

\end{thebibliography}\endgroup
